\documentclass[11pt,a4paper]{article}
\pdfoutput=1
\usepackage{a4wide}
\usepackage{amsmath}
\usepackage{amssymb}
\usepackage{amsfonts}
\usepackage{cite}
\usepackage{color}
\usepackage{adjustbox}
\usepackage{multirow}
\usepackage{pdflscape}
\usepackage{gensymb}
\usepackage{graphicx}
\usepackage{diagbox}
\usepackage{pifont}
\usepackage{bigstrut}
\usepackage[small,bf]{caption}
\usepackage{tabulary}
\usepackage{booktabs}
\usepackage{stackengine}

\newcommand{\be}{\begin{equation}}
\newcommand{\ee}{\end{equation}}

\def\1{\mathbf{1}}
\def\3{\mathbf{3}}
\def\2{\mathbf{2}}

\def\D{\Delta}

\def\th{\theta}

\def\gtap{\ \raisebox{-.4ex}{\rlap{$\sim$}} \raisebox{.4ex}{$>$}\ }
\def\ltap{\ \raisebox{-.4ex}{\rlap{$\sim$}} \raisebox{.4ex}{$<$}\ }

\allowdisplaybreaks
\numberwithin{equation}{section}
\makeatletter
\g@addto@macro\bfseries{\boldmath}
\makeatother

\begin{document}

\begin{titlepage}

\vspace*{-15mm}
\begin{flushright}
SISSA 15/2018/FISI \\ 
IPMU18-0058 \\
IPPP/18/22
\end{flushright}
\vspace*{8mm}

\begin{center}
{\bf\Large Assessing the Viability of $A_4$, $S_4$ and $A_5$ Flavour Symmetries}\\[2mm]
{\bf\Large for Description of Neutrino Mixing}\\[8mm]
S.~T.~Petcov$^{\,a,b,}$\footnote{Also at Institute of Nuclear Research and Nuclear Energy, Bulgarian Academy of Sciences, 1784 Sofia, Bulgaria.} 
and
A.~V.~Titov$^{\,c,}$\footnote{E-mail: \texttt{arsenii.titov@durham.ac.uk}} \\
\vspace{8mm}
$^{a}$\,{\it SISSA/INFN, Via Bonomea 265, 34136 Trieste, Italy} \\
\vspace{2mm}
$^{b}$\,{\it Kavli IPMU (WPI), University of Tokyo, 
5-1-5 Kashiwanoha, 277-8583 Kashiwa, Japan} \\
\vspace{2mm}
$^{c}$\,{\it Institute for Particle Physics Phenomenology, 
Department of Physics, Durham University,\\ 
South Road, Durham DH1 3LE, United Kingdom}
\end{center}
\vspace{8mm}

\begin{abstract}
\noindent 
We consider the $A_4$, $S_4$ and $A_5$ 
discrete lepton flavour symmetries
in the case of 3-neutrino mixing,
broken down to non-trivial residual symmetries in 
the charged lepton and neutrino sectors in such a way that 
at least one of them is a $Z_2$. 
Such symmetry breaking patterns lead to predictions 
for some of the three neutrino mixing angles and/or 
the leptonic Dirac CP violation phase $\delta$
of the neutrino mixing matrix. 
We assess the viability of these predictions by 
performing a statistical analysis 
which uses as an input the latest 
global data on the neutrino mixing parameters. 
We find 14 phenomenologically viable cases 
providing distinct predictions for some of the mixing angles 
and/or the Dirac phase $\delta$. 
Employing the current best fit values of the 
three neutrino mixing angles, 
we perform a statistical analysis of these cases 
taking into account the prospective uncertainties 
in the determination of the mixing angles, 
planned to be achieved in currently running (Daya Bay) 
and the next generation (JUNO, T2HK, DUNE) of neutrino oscillation experiments.
We find that only six cases would be compatible with these prospective data. 
We show that this number is likely to be further reduced 
by a precision measurement of $\delta$.
\end{abstract}
\end{titlepage}
\setcounter{footnote}{0}

\section{Introduction}
\label{sec:intro}

 Flavour is one of the biggest riddles in particle physics. 
In spite of the tremendous success of the Standard Theory, 
yet we do not know why the number of fermion generations is three, 
what determines the patterns of quark and lepton masses, 
and what the origins of quark and neutrino mixing are. 

  Since symmetries proved to be very powerful in guiding the laws 
of particle physics, it is natural to expect that 
symmetry might also be a clue to the 
solution of the flavour problem. 
For this reason, a variety of  \textit{flavour symmetries} 
have been proposed and explored 
in the attempts to understand the observed patterns 
of quark and/or neutrino mixing and of the 
quark and/or lepton masses.  
Symmetries described by both continuous groups, 
including $U(1)$, $SU(2)$, $U(2)$, $SU(3)$, $U(3)$ 
(see, e.g., \cite{Froggatt:1978nt,Barbieri:1996ae,Barbieri:1996ww,
Barbieri:1997tu,Petcov:1982ya,King:2003rf}), 
and discrete groups, such as $S_3$, $S_4$, $A_4$, $T'$, $A_5$, 
as well as the series $D_n$, $\Delta(3n^2)$, $\Delta(6n^2)$ 
with $n \in \mathbb{N}$ and $\Sigma$ groups 
(see, e.g., \cite{Altarelli:2010gt,Ishimori:2010au,King:2013eh} 
for reviews and original references) have been considered.
Discrete non-Abelian symmetries allow for rotations in the flavour space 
by fixed (large) angles, which is particularly attractive in view 
of the fact that two of the three neutrino mixing angles are large 
\cite{Esteban:2016qun,Capozzi:2017ipn,deSalas:2017kay}. 
Thus, neutrino mixing, as suggested, e.g., in \cite{Lam:2008rs},  
seems to be the appropriate flavour related structure 
to search for evidence of existence of an underlying flavour symmetry, 
and therefore for New Physics.

 In the framework of discrete flavour symmetry 
approach to 3-neutrino mixing~%
\footnote{For description of the reference 3-neutrino mixing scheme, 
see, e.g., \cite{Patrignani:2016xqp}.},
on which we will concentrate in the present article, 
it is assumed that at some high-energy scale there 
exists a (lepton) flavour symmetry 
described by a non-Abelian discrete (finite) group. 
The  lepton doublets of the 
three fermion generations are usually 
(but not universally)
assigned to an irreducible 3-dimensional representation 
of this group, because one aims to unify the three lepton flavours, 
and this is the case we will consider in the present article. 
At low energies the flavour symmetry has necessarily to be broken,
because the electron, muon and tauon 
charged leptons and the three massive neutrinos are distinct. 
Generally, the flavour symmetry group $G_f$ is broken in such a way that
the charged lepton and neutrino mass matrices, $M_e$ and  $M_\nu$~%
\footnote{More specifically, the charged lepton and neutrino mass matrices
of the charged lepton and neutrino Majorana (Dirac) mass terms 
written in left-right and right-left conventions, respectively.},
or more precisely, the combination $M_e M_e^\dagger$ and 
$M_\nu$ ($M^\dagger_\nu M_\nu$) in the Majorana 
(Dirac) neutrino case, are left invariant 
under the action of its Abelian subgroups 
$G_e$ and $G_\nu$, respectively.  
These \textit{residual symmetries} constrain the forms of 
the unitary matrices $U_e$ and $U_\nu$ diagonalising 
$M_e M_e^\dagger$ and $M_\nu$ ($M^\dagger_\nu M_\nu$), 
and thus of the Pontecorvo, Maki, Nakagawa, Sakata (PMNS)
neutrino mixing matrix $U_\mathrm{PMNS} = U_e^\dagger U_\nu$.

 If $G_e = Z_k$, $k > 2$ or $Z_m \times Z_n$, $m,n \geq 2$, and 
$G_\nu = Z_2 \times Z_2$ ($G_\nu = Z_k$, $k > 2$ or $Z_m \times Z_n$, 
$m,n \geq 2$) for Majorana (Dirac) neutrinos, the matrices $U_e$ and $U_\nu$ 
are fixed (up to permutations of columns and diagonal phase 
matrix on the right). This leads to certain fixed values of 
the solar, atmospheric and reactor 
neutrino mixing angles $\th_{12}$, $\th_{23}$ and $\th_{13}$ 
of the PMNS matrix~%
\footnote{Throughout this article we use the standard parametrisation 
of the PMNS matrix (see, e.g., \cite{Patrignani:2016xqp}).}.
Tri-bimaximal (TBM) mixing
\cite{Harrison:2002er,Harrison:2002kp,Xing:2002sw,He:2003rm} 
(see also \cite{Wolfenstein:1978uw}), characterised by 
$\th_{12} = \arcsin(1/\sqrt{3}) \approx 35\degree$, 
$\th_{23} = 45\degree$ and $\th_{13} = 0\degree$, 
is a well-known example of a symmetry form 
arising from a specific breaking pattern.
Namely, it can be naturally realised by breaking $G_f = S_4$ 
down to $G_e = Z_3$ and $G_\nu = Z_2 \times Z_2$ \cite{Lam:2008rs}. 
Other widely discussed examples include bimaximal (BM) mixing~%
\footnote{Bimaximal mixing can also be  
a consequence of the conservation of the lepton charge
$L' = L_e - L_{\mu} - L_{\tau}$ (LC) \cite{Petcov:1982ya}, 
supplemented by $\mu - \tau$ symmetry.}
($\th_{12} = \th_{23} = 45\degree$, $\th_{13} = 0\degree$) 
\cite{Vissani:1997pa,Barger:1998ta,Baltz:1998ey}, 
which can be derived from $G_f = S_4$ 
\cite{Altarelli:2009gn,Meloni:2011fx,Ding:2013eca}, and 
golden ratio A (GRA) mixing ($\th_{12} = \arctan(1/r) \approx 31\degree$, 
$\th_{23} = 45\degree$ and $\th_{13} = 0\degree$, 
$r=(1+\sqrt{5})/2$ being the golden ratio)
\cite{Datta:2003qg,Kajiyama:2007gx}, 
which can be obtained 
breaking $G_f = A_5$ to $G_e = Z_5$ and $G_\nu = Z_2 \times Z_2$ 
\cite{Everett:2008et,Ding:2011cm}. 
All these highly symmetric mixing patterns, however,
were ruled out once $\th_{13}$ was measured 
and found to have a non-zero value, $\theta_{13}\cong 0.15$. 
The fact that $\theta_{13}$ turned out to have a relatively large value
opened up a possibility of 
establishing the status of Dirac CP violation (CPV) in the lepton sector  
by measuring the Dirac phase $\delta$ present in the PMNS matrix. 
At the same time it implied, in particular, that 
the TBM, BM (LC), GRA and other symmetry forms of the PMNS matrix 
predicting $\theta_{13} = 0$~%
\footnote{Additional examples of symmetry forms predicting $\theta_{13} = 0$ 
include the golden ratio B (GRB) form 
($\th_{12} = \arccos(r/2) = 36\degree$, $\th_{23} = 45\degree$) 
\cite{Rodejohann:2008ir,Adulpravitchai:2009bg}
and the hexagonal (HG) form 
($\th_{12} = 30\degree$, $\th_{23} = 45\degree$)
\cite{Albright:2010ap,Kim:2010zub}.}
have to be ``perturbed'',
so that $\theta_{13}$, as well as $\theta_{12}$ and $\theta_{23}$,
have values compatible with the experimentally determined values.
When, for example, the requisite ``perturbations'' are provided by the matrix 
$U_e$ and have the simple form of
a $U(2)$ transformation in a plane or a product of two
$U(2)$ transformations each in a plane, 
the cosine of the phase $\delta$ was shown 
\cite{Petcov:2014laa,Girardi:2015vha}
to satisfy a \textit{sum rule} by which it is expressed in terms 
of the three neutrino mixing angles and an angle parameter 
which takes discrete values depending on the 
underlying symmetry form (TBM, BM (LC), GRA, GRB, HG) of the PMNS matrix.
Analogous sum rule for $\cos\delta$ arises when, e.g., 
the TBM symmetry form of $U_\mathrm{PMNS}$ is ``perturbed'' 
on the right by a matrix describing a $U(2)$ transformation in the 
1-3 plane \cite{Grimus:2008tt} or 2-3 plane \cite{Albright:2008rp}~%
\footnote{These two sum rules can be obtained from the general 
results derived in  \cite{Girardi:2015rwa}.} 
(see, e.g., \cite{Petcov:2017ggy} for a recent review of 
the discussed sum rules). The measurement of $\theta_{13}\cong 0.15$
gave also a boost to investigating alternative
flavour symmetry breaking patterns 
in attempt to explain the special 
structure of the PMNS matrix.

 In \cite{Girardi:2015rwa} all symmetry breaking patterns, 
i.e., all possible combinations 
of residual symmetries, which could lead to 
correlations between some of  
the three neutrino mixing angles and/or between the neutrino mixing angles 
and the Dirac CPV phase $\delta$, were considered.
Namely, 
(A) $G_e = Z_2$ and $G_{\nu} = Z_k$, $k > 2$ or $Z_m \times Z_n$, $m,n \geq 2$; 
(B) $G_e = Z_k$, $k > 2$ or $Z_m \times Z_n$, $m,n \geq 2$ and $G_{\nu} = Z_2$;
(C) $G_e = Z_2$ and $G_{\nu} = Z_2$;
(D) $G_e$ is fully broken and $G_{\nu} = Z_k$, $k > 2$ or $Z_m \times Z_n$, $m,n \geq 2$; 
and (E) $G_e = Z_k$, $k > 2$ or $Z_m \times Z_n$, $m,n \geq 2$ and $G_{\nu}$ is fully broken.
For each pattern, sum rules, i.e., 
relations between the neutrino mixing angles and/or between 
the neutrino mixing angles and the Dirac CPV phase $\delta$, when present,
were derived.
Neutrino mixing sum rules can be present also 
in the case of pattern D (E) if due to additional
assumptions (e.g., additional symmetries)
the otherwise unconstrained unitary
matrix $U_e$ ($U_\nu$) is
constrained to have the specific form of
a matrix of $U(2)$ transformation in a plane or
of the product of two $U(2)$ transformations
in two different planes
\cite{Petcov:2014laa,Girardi:2015vha,Girardi:2015rwa,Marzocca:2013cr,Girardi:2014faa}.
Therefore, the cases of patterns D and E
leading to interesting phenomenological predictions
are ``non-minimal'' from the point
of view of the symmetries employed
(see, e.g., \cite{Meroni:2012ty,Hagedorn:2012pg,Antusch:2013kna,Chen:2013wba,
Girardi:2013sza,Gehrlein:2014wda}), 
compared to patterns A, B and C characterised by 
non-trivial residual symmetries present in both 
charged lepton and neutrino sectors, which originate
from just one non-Abelian flavour symmetry.

 In the present article, we concentrate on patterns A, B and C, 
 assuming $G_f = A_4~(T^\prime)$, $S_4$ and $A_5$. 
When choosing these flavour symmetries, we are guided by minimality: 
$A_4~(T^\prime)$, $S_4$ and $A_5$ are among smallest
(in terms of the number of elements) 
discrete groups admitting a 3-dimensional irreducible representation.
In \cite{Girardi:2015rwa} predictions for the 
mixing angles and $\cos\delta$ 
have been obtained in the cases of patterns A, B and C
originating from $G_f = A_4~(T^\prime)$~%
\footnote{The results obtained in \cite{Girardi:2015rwa}
and in the present article for the group $A_4$ are valid also
for $T^\prime$, since when working
with the 3-dimensional and 1-dimensional irreducible representations,
$T^\prime$ and  $A_4$ lead to the same results \cite{Feruglio:2007uu}.}, 
$S_4$ and $A_5$, using
the best fit values
of other (free) mixing angles entering into the sum rules of interest. 
In this work, we perform a statistical analysis of the sum rule predictions 
derived in  \cite{Girardi:2015rwa}, 
taking into account 
(i)~the latest global data on the neutrino
mixing parameters \cite{NuFITv32Jan2018}, 
and (ii)~the prospective uncertainties in the determination of
the neutrino mixing angles, which are planned to be achieved in
the next generation of neutrino oscillation experiments.
The results of this analysis clearly demonstrate how phenomenologically viable
the considered cases, and hence the $A_4$, $S_4$ and $A_5$ 
flavour symmetries, are. 

 The layout of the remainder of this article is as follows. 
In Section~\ref{sec:patterns}, we recall the framework and recapitulate  
the relevant sum rules derived in \cite{Girardi:2015rwa}. 
In Section~\ref{sec:groups}, we give a brief description of 
the discrete groups $A_4$, $S_4$ and $A_5$ emphasising 
the features relevant for our analysis. 
In Section~\ref{sec:predictions}, we study in detail the predictions 
for the neutrino mixing angles and the Dirac CPV phase. 
We perform a statistical analysis of the predictions for 
$\sin^2\th_{12}$, $\sin^2\th_{23}$ and $\cos\delta$ 
taking into account first the current and then 
the prospective uncertainties in the determination of the mixing parameters.
Finally, we summarise the obtained results and conclude 
in Section~\ref{sec:conclusions}.

\section{Residual Symmetry Patterns and Sum Rules}
\label{sec:patterns}

 In this section, we briefly summarise the results for patterns A, B and C 
obtained in ref.~\cite{Girardi:2015rwa}. We will use these results 
in Section~\ref{sec:predictions} 
to perform a statistical analysis of the predictions 
for the mixing angles and $\cos\delta$.

 \textit{Pattern A: $G_e = Z_2$ and $G_{\nu} = Z_k$, $k > 2$ 
or $Z_m \times Z_n$, $m,n \geq 2$.}
The $Z_2$ residual symmetry in the charged lepton sector fixes the matrix $U_e$ 
up to a $U(2)$ transformation in the $i$-$j$ plane. 
This transformation can be parametrised in terms of a matrix containing 
one angle and three phases. Two of the three phases 
can be removed by a redefinition of the charged lepton fields. 
Therefore the three neutrino mixing angles and the Dirac phase 
are expressed in terms of the remaining two free parameters. 
As a result, correlations between the observables arise.
Namely, the considered type of residual symmetries leads to 
sum rules for $\sin^2\th_{23}$ and $\cos\delta$,
except in one case (case A3, see further) in which 
$\sin^2\th_{12}$ and $\sin^2\th_{13}$ are predicted and 
$\delta$ is not constrained.

 Depending on the plane in which the $U(2)$ transformation is performed, 
one has three cases. The first one, which we denote as A1, 
corresponds to the transformation 
in the 1-2 plane and leads to the following sum rules:
\be
\sin^2 \theta_{23} = 1 -
\frac{\cos^2 \theta^{\circ}_{13} \cos^2 \theta^{\circ}_{23}}{1-\sin^2\th_{13}}\,,
\label{eq:ss23A1}
\ee
\be
\cos \delta = \frac{\cos^2 \theta_{13} (\sin^2 \theta^{\circ}_{23} - \cos^2 \theta_{12}) + \cos^2 \theta^{\circ}_{13} \cos^2 \theta^{\circ}_{23} (\cos^2 \theta_{12} - \sin^2 \theta_{12} \sin^2 \theta_{13})}{\sin 2 \theta_{12} \sin \theta_{13} |\cos \theta^{\circ}_{13} \cos \theta^{\circ}_{23}| (\cos^2 \theta_{13} - \cos^2 \theta^{\circ}_{13} \cos^2 \theta^{\circ}_{23})^{\frac{1}{2}}}\,,
\label{eq:cosdeltaA1}
\ee
%
where the angles $\th^\circ_{13}$ and $\th^\circ_{23}$ are fixed 
once the flavour symmetry group $G_f$ and 
the residual symmetry subgroups $G_e$ and $G_\nu$ are specified.
In the second case, A2, which corresponds to the free $U(2)$ transformation 
in the $1$-$3$ plane, one has different relations: 
\be
\sin^2 \theta_{23} = \frac{\sin^2 \theta^{\circ}_{23}}{1 - \sin^2 \theta_{13}}\,, 
\label{eq:ss23A2}
\ee
\be
\cos \delta = -\frac{\cos^2 \theta_{13} (\cos^2 \theta^{\circ}_{12} \cos^2 \theta^{\circ}_{23} - \cos^2 \theta_{12}) + \sin^2 \theta^{\circ}_{23} (\cos^2 \theta_{12} - \sin^2 \theta_{12} \sin^2 \theta_{13})}{\sin 2 \theta_{12} \sin \theta_{13} |\sin \theta^{\circ}_{23}| (\cos^2 \theta_{13} - \sin^2 \theta^{\circ}_{23})^{\frac{1}{2}}}\,,
\label{eq:cosdeltaA2}
\ee
%
where also the angle $\theta^{\circ}_{12}$ is  fixed 
once $G_f$, $G_e$ and $G_\nu$ are specified.
Finally, case A3 corresponding to the $U(2)$ transformation 
in the $2$-$3$ plane predicts 
$\sin^2\th_{13} = \sin^2\th^\circ_{13}$ and $\sin^2\th_{12} = \sin^2\th^\circ_{12}$, 
while $\cos\delta$ remains unconstrained.

 \textit{Pattern B: $G_e = Z_k$, $k > 2$ or $Z_m \times Z_n$, $m,n \geq 2$ 
and $G_{\nu} = Z_2$.} 
The residual $Z_2$ symmetry determines the matrix $U_\nu$ 
up to a $U(2)$ transformation in the $i$-$j$ plane. For Dirac neutrinos, 
two of the three phases parametrising this transformation can be removed 
by a re-phasing of the neutrino fields. For Majorana neutrinos, this is not possible, 
and these two phases will contribute to the Majorana phases in the PMNS matrix.
In either case, they will not enter into the expressions 
for the mixing angles and the Dirac phase, 
which depend on the remaining two free parameters (an angle and a phase).
Pattern B leads to sum rules for $\sin^2\th_{12}$ and $\cos\delta$, 
again except in one case (case B3, see further) in which 
$\sin^2\th_{23}$ and $\sin^2\th_{13}$ are predicted and 
$\delta$ is not constrained.

 Again, depending on the plain of the $U(2)$ transformation, 
we have three cases. 
Case B1 corresponding to $(ij) = (13)$ yields 
\be
\sin^2 \theta_{12} = \frac{\sin^2 \theta^{\circ}_{12}}{1 - \sin^2 \theta_{13}}\,,
\label{eq:ss12B1}
\ee
\be
\cos \delta = -\frac{\cos^2 \theta_{13} (\cos^2 \theta^{\circ}_{12} \cos^2 \theta^{\circ}_{23} - \cos^2 \theta_{23}) + \sin^2 \theta^{\circ}_{12} (\cos^2 \theta_{23} - \sin^2 \theta_{13} \sin^2 \theta_{23})}
{\sin 2 \theta_{23} \sin \theta_{13} |\sin \theta^{\circ}_{12}| (\cos^2 \theta_{13} - \sin^2 \theta^{\circ}_{12})^{\frac{1}{2}}}\,,
\label{eq:cosdeltaB1}
\ee
%
where $\th^\circ_{12}$ and $\th^\circ_{23}$ are fixed once the symmetries are specified.
In case B2, $(ij) = (23)$, the sum rules of interest read: 
\be
\sin^2 \theta_{12} = 1 - 
\frac{\cos^2\theta^{\circ}_{12} \cos^2\theta^{\circ}_{13}  }{1 - \sin^2\theta_{13}}\,,
\label{eq:ss12B2}
\ee
\be
\cos \delta = 
\frac{\cos^2 \theta_{13} (\sin^2 \theta^{\circ}_{12} - \cos^2 \theta_{23}) + \cos^2 \theta^{\circ}_{12} \cos^2 \theta^{\circ}_{13} ( \cos^2 \theta_{23} - \sin^2 \theta_{13} \sin^2 \theta_{23} )}
{ \sin 2 \theta_{23} \sin \theta_{13} | \cos \theta^{\circ}_{12} \cos \theta^{\circ}_{13}| (\cos^2 \theta_{13} - \cos^2 \theta^{\circ}_{12} \cos^2 \theta^{\circ}_{13} )^{\frac{1}{2}}}\,.
\label{eq:cosdeltaB2}
\ee
%
At last, case B3, $(ij) = (12)$, leads to 
$\sin^2\th_{13} = \sin^2\th^\circ_{13}$ and $\sin^2\th_{23} = \sin^2\th^\circ_{23}$, 
and no sum rule for $\cos\delta$.

 \textit{Pattern C: $G_e = Z_2$ and $G_{\nu} = Z_2$.} 
In this case, both $U_e$ and $U_\nu$ are determined up to $U(2)$ 
transformations in the $i$-$j$ and $k$-$l$ planes, respectively.
Thus, we have four free parameters (two angles and two phases) 
in terms of which $\th_{ij}$ and $\delta$ are expressed.
However, as shown in \cite{Girardi:2015rwa},
this number is reduced to three after
an appropriate rearrangement of these parameters.
As a consequence, a sum rule for either $\cos\delta$ or one of 
$\sin^2\th_{ij}$ arises.

 Depending on the planes in which the free $U(2)$ transformations 
are performed, we have nine possibilities. 
We number them as in \cite{Girardi:2015rwa}, i.e., cases C1--C9.
Four of them lead to sum rules for $\cos\delta$, which we summarise bellow.
\begin{align}
&\text{C1, $(ij,kl) = (12,13)$:} \quad
\cos \delta = 
\dfrac{\sin^2 \theta^{\circ}_{23} - \cos^2 \theta_{12} \sin^2 \theta_{23} - \cos^2 \theta_{23} \sin^2 \theta_{12} \sin^2 \theta_{13}}
{\sin \theta_{13} \sin 2 \theta_{23} \sin \theta_{12} \cos \theta_{12}}\,, 
\label{eq:cosdeltaC1}\\[0.2cm]
&\text{C3, $(ij,kl) = (12,23)$:} \quad
\cos \delta = 
\dfrac{\sin^2 \theta_{12} \sin^2 \theta_{23} - \sin^2 \theta^{\circ}_{13} + \cos^2 \theta_{12} \cos^2 \theta_{23} \sin^2 \theta_{13}}
{\sin \theta_{13} \sin 2 \theta_{23} \sin \theta_{12} \cos \theta_{12}}\,, \\[0.2cm]
&\text{C4, $(ij,kl) = (13,23)$:} \quad 
\cos \delta = 
\dfrac{\sin^2 \theta^{\circ}_{12} - \cos^2 \theta_{23} \sin^2 \theta_{12} - \cos^2 \theta_{12} \sin^2 \theta_{13} \sin^2 \theta_{23}}
{\sin \theta_{13} \sin 2 \theta_{23} \sin \theta_{12} \cos \theta_{12}}\,, \\[0.2cm]
&\text{C8, $(ij,kl) = (13,13)$:} \quad 
\cos \delta = 
\dfrac{\cos^2 \theta_{12} \cos^2 \theta_{23} - \cos^2 \theta^{\circ}_{23} + 
\sin^2 \theta_{12} \sin^2 \theta_{23} \sin^2 \theta_{13}}
{\sin \theta_{13} \sin 2 \theta_{23} \sin \theta_{12} \cos \theta_{12}}\,.
\label{eq:cosdeltaC8}
\end{align}
%
The neutrino mixing angles in these cases can be treated as free parameters. 
Other two cases, C5 and C9, yield correlations between 
$\sin^2\th_{12}$ and $\sin^2\th_{13}$. Namely, 
\begin{align}
&\text{C5, $(ij,kl) = (23,13)$:} \quad 
\sin^2 \theta_{12} = \frac{\sin^2 \theta^{\circ}_{12}}{1 - \sin^2 \theta_{13}}\,, \\[0.2cm]
&\text{C9, $(ij,kl) = (23,23)$:} \quad
\sin^2 \theta_{12} = \frac{\sin^2 \theta^\circ_{12} - \sin^2 \theta_{13}}{1 - \sin^2 \theta_{13}}\,.
\label{eq:ss12C9}
\end{align}
%
In cases C2 and C7, instead, there are correlations between
$\sin^2\th_{23}$ and $\sin^2\th_{13}$:
\begin{align}
&\text{C2, $(ij,kl) = (13,12)$:} \quad 
\sin^2 \theta_{23} = \frac{\sin^2 \theta^{\circ}_{23}}{1 - \sin^2 \theta_{13}}\,, 
\label{eq:ss23C2}\\[0.2cm]
&\text{C7, $(ij,kl) = (12,12)$:} \quad 
\sin^2 \theta_{23} = \frac{\sin^2 \theta^\circ_{23} - \sin^2 \theta_{13}}
{1 - \sin^2 \theta_{13}}\,.
\label{eq:ss23C7}
\end{align}
%
Finally, in case C6, $(ij,kl) = (23,12)$, $\sin^2\th_{13}$ 
is predicted to be equal to $\sin^2\th^\circ_{13}$. 
In cases C2, C5, C6, C7 and C9, $\cos\delta$ remains unconstrained.

 In Section~\ref{sec:predictions}, we will apply these sum rules 
to derive predictions from the $A_4$, $S_4$ and $A_5$ flavour symmetries. 
We recall that the parameters $\th^\circ_{ij}$ are fixed once 
the flavour symmetry group and the residual symmetry subgroups are specified.

\section{The $A_4$, $S_4$ and $A_5$ Symmetries}
\label{sec:groups}

 The alternating group $A_4$ is the group of even permutations on four objects. 
It is isomorphic to the group of rotational symmetries of a regular tetrahedron.  
All its twelve elements can be expressed in terms of two generators, 
usually denoted as $S$ and $T$, which satisfy the following presentation rules: 
\be
S^2 = T^3 = (ST)^3 = E,   
\ee
%
$E$ being the identity of the group.
$A_4$ possesses four irreducible representations: 
three 1-dimensional and one 3-dimensional. 
The eight Abelian subgroups of $A_4$ amount to  
three $Z_2$, four $Z_3$ and one Klein group $K_4$ isomorphic to $Z_2\times Z_2$.
The detailed list of them can be found in \cite{Tanimoto:2015nfa}.
All these subgroups can serve as residual symmetries 
of the charged lepton and neutrino mass matrices~%
\footnote{We recall that in the case of Majorana neutrinos 
the residual symmetry $G_\nu$ 
can be either $Z_2$ or $Z_2\times Z_2$.}.
In the case of $A_4$, we have pairs 
$(G_e,G_\nu) = (Z_2,Z_3)$ and $(Z_2,Z_2\times Z_2)$
corresponding to pattern A of residual symmetries, 
$ (Z_3,Z_2)$ and $(Z_2\times Z_2,Z_2)$ to pattern B, 
and $(Z_2,Z_2)$ to pattern C.

 The symmetric group $S_4$ is the group of all permutations on four objects. 
It is isomorphic to the group of rotational symmetries of a cube. 
It contains $A_4$ as a subgroup. 
The 24 elements of $S_4$ can be 
generated by two transformations $\tilde S$ and $\tilde T$ 
(see, e.g.,\cite{Altarelli:2010gt,Ishimori:2010au}). 
However, in the context of non-Abelian discrete symmetry approach 
to neutrino mixing, it often 
proves convenient to use the three generators 
 $S$, $T$ and $U$, satisfying~%
\footnote{This presentation of $S_4$ is convenient, because 
$S$ and $T$ alone generate the $A_4$ subgroup of $S_4$.}
the following presentation rules:
\be
S^2 = T^3 = U^2 = (ST)^3 = (SU)^2 = (TU)^2 = (STU)^4 = E\,.
\ee
%
The results from \cite{Girardi:2015rwa} we are going to use in 
what follows were obtained working with the three generators $S$, $T$
and $U$ of $S_4$.   
The group admits five irreducible
representations: two singlet, one doublet 
and two triplet. The list of 20 Abelian subgroups of $S_4$ consists of 
nine $Z_2$, four $Z_3$, three $Z_4$ and four $Z_2\times Z_2$ groups 
(see, e.g., \cite{Tanimoto:2015nfa}).

 The alternating group $A_5$ is the group of even permutations on 
five objects. It is isomorphic to the group of rotational symmetries 
of a regular icosahedron. Obviously, $A_4$ is contained in $A_5$ as a subgroup.
The 60 elements of $A_5$ can be defined in terms of two generators 
$S$ and $T$, satisfying~%
\footnote{We note that the generators $S$ and $T$ of $A_5$ are different 
from the corresponding generators of $A_4$ and $S_4$ denoted by the same 
letters.}
\be
S^2 = T^5 = (ST)^3 = E\,.
\ee
%
In addition to the two 3-dimensional irreducible representations, the group 
possesses one singlet, one 4-dimensional and one 5-dimensional representations.
In total, $A_5$ has 36 Abelian subgroups: fifteen $Z_2$, ten $Z_3$, 
five $Z_2\times Z_2$ and six $Z_5$. The complete list of them 
can be found in \cite{Ding:2011cm}.

 In \cite{Girardi:2015rwa} all possible pairs of the 
Abelian subgroups of $A_4$, $S_4$ and $A_5$ listed above, 
which correspond to patterns A, B and C discussed in the previous section, 
have been considered. 
Using the suitable parametrisation of the PMNS matrix in each case, 
we have obtained the values of the fixed parameters $\sin^2\th^\circ_{ij}$  
relevant for the sum rules given in eqs.~\eqref{eq:ss23A1}--\eqref{eq:ss23C7}. 
Finally, employing these sum rules and the best fit values of the neutrino mixing angles, 
we have derived predictions for $\cos\delta$ and $\sin^2\th_{ij}$. 
They are summarised in Tables~9--11 in \cite{Girardi:2015rwa}.

In the next section, we first update the predictions
for $\cos\delta$ and $\sin^2\th_{ij}$ 
using the best fit values of the mixing angles obtained
in the latest global analysis 
of neutrino oscillation data \cite{NuFITv32Jan2018}.
Secondly, and most importantly, we perform a statistical
analysis of the sum rule predictions, taking into account 
(i) the latest global data on the neutrino mixing
parameters \cite{NuFITv32Jan2018}, 
and (ii) the prospective uncertainties in the determination
of the mixing angles, which are planned to be achieved in the
next generation of neutrino oscillation experiments.
As we will see, the results of our analysis clearly demonstrate how 
phenomenologically viable the cases under consideration 
are at the moment and what the perspective for testing them is.

\section{Predictions for the Mixing Angles and the Dirac CPV Phase}
\label{sec:predictions}

 Before proceeding to the numerical results, we would like to make a comment on 
the number of possible cases we have, since a priori this number is large,  
and one could be surprised by a relatively small number of
\textit{viable} cases we find and present in what follows.

 Let us consider as an example $G_f = A_4$.  
First, we examine the residual symmetries $G_e$ and $G_\nu$,
which lead to fully specified mixing patterns. 
There are four such types of pairs $(G_e,G_\nu)$. We comment on each of them below.
\begin{itemize} 
\item $(G_e,G_\nu) = (Z_2\times Z_2,Z_2\times Z_2)$.
In this case, the matrices $U_e$ and $U_\nu$ are the same 
(up to permutations of columns and diagonal phase matrices on the right).  
Therefore, the PMNS matrix is given by the unit matrix 
up to permutations of rows and columns 
and possible Majorana phases. This case is clearly non-viable.
\item $(G_e,G_\nu) = (Z_3,Z_2\times Z_2)$. 
There are four such pairs in the case of $G_f = A_4$. 
All of them are conjugate to each other. 
As is well known, 
two pairs of residual symmetries, which are conjugate to each other 
under an element of $G_f$, lead to the same PMNS matrix 
(see, e.g., \cite{Feruglio:2012cw,Ding:2013bpa}). 
Thus, it is enough to consider only one of them.
The resulting PMNS matrix is fixed up to permutations of rows and columns, 
but is not viable (see, e.g., \cite{Ding:2013bpa}).
\item $(G_e,G_\nu) = (Z_2\times Z_2,Z_3)$. 
Again, four possible pairs are conjugate to each other and lead to
the same PMNS fixed up to permutations of rows and columns. 
This case is not consistent with the data either.
\item $(G_e,G_\nu) = (Z_3,Z_3)$. The 16 possible $(Z_3^{g_e},Z_3^{g_\nu})$ pairs, 
$g_e$ and $g_\nu$ being the generating elements of
the $Z_3^{g_e}$ and $Z_3^{g_\nu}$ subgroups, respectively, 
fall into two groups. 
There are four pairs with $g_e = g_\nu$ 
and twelve pairs with $g_e \neq g_\nu$. 
The former four are conjugate to each other and lead to the same PMNS matrix, 
which corresponds to the unit matrix up to permutations of rows and columns.
The latter twelve are also related to each other by a similarity transformation, 
thus leading to the same PMNS matrix fixed, as always, up to permutations of 
rows and columns. This pattern is not viable as well.
\end{itemize}
%
Secondly, considering patterns A, B and C 
of the residual symmetries $G_e$ and $G_\nu$, 
which do not lead to fully specified $U_\mathrm{PMNS}$, 
we have five possibilities.
\begin{itemize}
\item $(G_e,G_\nu) = (Z_2,Z_2\times Z_2)$. 
There are three such pairs for $G_f = A_4$, 
all of them being conjugate to each other. 
Thus, it is enough to consider only one of them. 
However, in the case of $A_4$, any $Z_2$ is a subgroup of the $Z_2\times Z_2$. 
As shown in \cite{Girardi:2015rwa}, 
$(G_e,G_\nu) = (Z_2^{g_e},Z_2^{g_\nu}\times Z_2)$ and 
$(G_e,G_\nu) = (Z_2^{g_e}\times Z_2,Z_2^{g_\nu})$ with 
$g_e$ and $g_\nu$ belonging to the same 
$Z_2\times Z_2$ subgroup of $G_f$, lead to some entries 
of $U_\mathrm{PMNS}$ being zero, which is ruled out
by the data \cite{NuFITv32Jan2018}.
\item $(G_e,G_\nu) = (Z_2,Z_3)$. 
One can demonstrate that the twelve possible pairs are all conjugate to each other, 
and thus, they predict the same PMNS matrix.
The latter is defined up to a free $U(2)$ transformation  
applied from the left in the $i$-$j$ plane (3 possibilities)  
as explained in Section~\ref{sec:patterns}, and up to permutations of columns.
\item $(G_e,G_\nu) = (Z_2\times Z_2,Z_2)$. 
There are three  such pairs, 
all of them being related to each other by a similarity transformation. 
The same argument as for $(G_e,G_\nu) = (Z_2,Z_2\times Z_2)$ works in this case. 
The resulting PMNS matrix is not viable, because it contains zero entries.
\item $(G_e,G_\nu) = (Z_3,Z_2)$. 
The twelve possible pairs are all conjugate to each other, 
and thus, they predict the same PMNS matrix.
It is defined up to a free $U(2)$ transformation 
applied from the right in the $i$-$j$ plane (3 possibilities)  
as explained in Section~\ref{sec:patterns}, and up to permutations of columns. 
As we will see, the case of the transformation in the 1-3 plane is the only case 
consistent with the data.
\item $(G_e,G_\nu) = (Z_2,Z_2)$. 
The nine possible $(Z_2^{g_e},Z_2^{g_\nu})$ pairs can be partitioned 
into two equivalent classes. 
The first class contains three pairs with $g_e = g_\nu$, which are conjugate to each other. 
They lead to the same PMNS matrix with zero entries 
(see, e.g., \cite{Girardi:2015rwa,Penedo:2017vtf}). 
The second class consists of six pairs with $g_e \neq g_\nu$, all of them being related to
each other by a similarity transformation. 
Since $g_e,\,g_\nu \in Z_2\times Z_2 \subset A_4$, 
the resulting PMNS matrix contains a zero entry in this case as well  \cite{Girardi:2015rwa,Penedo:2017vtf}. 
Therefore, the considered pattern is not viable.
\end{itemize}

 Thus, the total number of cases is 64 
(up to permutations of rows and columns of the PMNS matrix). 
Of these only 8 lead to distinct predictions for $U_\mathrm{PMNS}$, 
while only five cases a priori can be phenomenologically viable~%
\footnote{By ``a priori'' we mean that they lead to $U_\mathrm{PMNS}$ 
without zero entries.}. 
Similar analyses can be performed for the $S_4$ and $A_5$ symmetries. 

 In our further analysis we require that all three mixing
angles lie simultaneously in their respective $3\sigma$ ranges,
and that the sum rule for $\cos\delta$,
whenever present, leads to $|\cos\delta| \leq 1$ (see further).
Thus, the number of the remaining cases gets further reduced
by these requirements.

\subsection{Analysis with Best Fit Values}
\label{subsec:bfv}

 In this subsection, we use the best fit values of the mixing angles 
found in the latest global analysis \cite{NuFITv32Jan2018} to 
update the numerical predictions for $\cos\delta$ and $\sin^2\th_{ij}$ 
obtained in \cite{Girardi:2015rwa}. For convenience, we present 
the current best fit values of $\sin^2\th_{ij}$ and $\delta$ 
along with their respective $3\sigma$ ranges in Table~\ref{tab:parameters}.
\begin{table}
\centering
\renewcommand*{\arraystretch}{1.2}
\begin{tabular}{lll} 
\toprule
Parameter & Best fit & $3\sigma$ range \\ 
\midrule
$\sin^2\th_{12}$ & $0.307$ & $0.272 - 0.346$ \\[0.2cm]
$\sin^2\th_{23}$ (NO) & $0.538$ & $0.418 - 0.613$ \\
$\sin^2\th_{23}$ (IO) & $0.554$ & $0.435 - 0.616$ \\[0.2cm]
$\sin^2\th_{13}$ (NO) & $0.02206$ & $0.01981 - 0.02436$ \\
$\sin^2\th_{13}$ (IO) & $0.02227$ & $0.02006 - 0.02452$ \\[0.2cm]
$\delta$ [\degree] (NO) & $234$ & $144 - 374$ \\
$\delta$ [\degree] (IO) & $278$ & $192 - 354$ \\
\bottomrule
\end{tabular}
\caption{The best fit values and 3$\sigma$ ranges of the 
neutrino mixing parameters obtained in the latest global 
analysis of neutrino oscillation data \cite{NuFITv32Jan2018}. 
NO (IO) stands for normal (inverted) ordering of the neutrino mass spectrum.}
\label{tab:parameters}
\end{table}

 In the case of $G_f = A_4$, there is \textit{only one}
phenomenologically viable case. 
Namely, this is case B1 with $(G_e,G_\nu) = (Z_3,Z_2)$, which yields 
$(\sin^2\th^\circ_{12},\sin^2\th^\circ_{23}) = (1/3,1/2)$ and corresponds to 
the TBM mixing matrix corrected from the right by a $U(2)$
transformation in the 1-3 plane.
Making use of eqs.~\eqref{eq:ss12B1} and \eqref{eq:cosdeltaB1} 
and the current best fit values of the mixing angles 
for the NO neutrino mass spectrum, we find 
the predictions summarised in Table~\ref{tab:A4}.
\begin{table}
\centering
\renewcommand*{\arraystretch}{1.2}
\begin{tabular}{lllll}
\toprule
$(G_e,G_\nu)$ & Case & $\sin^2\theta^{\circ}_{ij}$ & $\cos \delta$ & $\sin^2 \theta_{ij}$ \\
\midrule
$(Z_3,Z_2)$ & B1 & $(\sin^2 \theta^{\circ}_{12},\sin^2 \theta^{\circ}_{23}) = (1/3,1/2)$ & $-0.353$ & $\sin^2 \theta_{12} = 0.341$ \\
\bottomrule
\end{tabular}
\caption{The only viable case for $G_f = A_4$.
The values of $\cos \delta$ and $\sin^2 \theta_{12}$ 
are obtained using the best fit values of 
$\sin^2 \theta_{13}$ and $\sin^2 \theta_{23}$ for NO.}
\label{tab:A4}
\end{table}
%
In the next subsections we will investigate in detail how these
predictions modify, if one takes into account the uncertainties
in the determination of the neutrino mixing parameters.

 In the case of $G_f = S_4$, the number of viable cases is larger, 
namely, there are eight viable cases.
We summarise them in Table~\ref{tab:S4}. 
\begin{table}[t]
\centering
\renewcommand*{\arraystretch}{1.2}
\begin{tabular}{lllll}
\toprule
$(G_e,G_\nu)$ & Case & $\sin^2\th^\circ_{ij}$ & $\cos \delta$ & $\sin^2 \theta_{ij}$\\
\midrule
\multirow{2}{*}{$(Z_3,Z_2)$} & B1 & $(\sin^2 \theta^{\circ}_{12},\sin^2 \theta^{\circ}_{23}) = (1/3,1/2)$ & 
$-0.353$ & $\sin^2 \theta_{12} = 0.341$ \\   
& B2S$_4$ & $(\sin^2 \theta^{\circ}_{12},\sin^2 \theta^{\circ}_{13}) = (1/6,1/5)$ &
$0.167$ & $\sin^2 \theta_{12} = 0.318$ \\  
\midrule
\multirow{6}{*}{$(Z_2,Z_2)$} & C1 & $\sin^2 \theta^{\circ}_{23} = 1/4$ & $-1^*$ & not fixed\\   
& C2S$_4$ & $\sin^2 \theta^{\circ}_{23} = 1/2$ & not fixed &$\sin^2 \theta_{23} = 0.511$ \\    
& C3 & $\sin^2 \theta^{\circ}_{13} = 1/4$ & $-1^*$ & not fixed\\   
& C4 & $\sin^2 \theta^{\circ}_{12} = 1/4$ & $1^*$ & not fixed \\
& C7S$_4$ & $\sin^2 \theta^{\circ}_{23} = 1/2$ & not fixed & $\sin^2 \theta_{23} = 0.489$ \\
& C8 & $\sin^2 \theta^{\circ}_{23} = 3/4$ & $1^*$ & not fixed \\
\bottomrule
\end{tabular}
\caption{The viable cases for $G_f = S_4$.
The values of $\cos \delta$ and $\sin^2 \theta_{12} / \sin^2 \theta_{23}$ 
are obtained using the best fit values of the relevant (not fixed) mixing angles for NO.
In the cases marked with an asterisk, physical values of $\cos \delta$  
cannot be obtained employing the best fit values of the mixing angles, 
but they are achievable fixing two angles to their best fit values and 
varying the third one in its $3\sigma$ range.}
\label{tab:S4}
\end{table}
%
In the cases marked with an asterisk, the use of the best fit values
of the mixing angles leads to unphysical values of
$\cos\delta$, i.e., $|\cos\delta|>1$, 
which reflects the fact that these cases cannot provide a good description 
of the best fit values of all three mixing angles simultaneously. 
However, the physical values of $\cos\delta$ can be obtained in these cases 
fixing two angles to their best fit values and varying the third one 
in its $3\sigma$ range. 

 Finally, for $G_f = A_5$, requiring the compatibility with the data 
in the way explained above, we find 13 viable cases. 
They are presented in Table~\ref{tab:A5}. 
\begin{table}
\centering
\renewcommand*{\arraystretch}{1.2}
\begin{adjustbox}{max width=\textwidth}
\begin{tabular}{lllll}
\toprule
$(G_e,G_\nu)$ & Case & $\sin^2\th^\circ_{ij}$ & $\cos\delta$ & $\sin^2 \theta_{ij}$\\
\midrule
\multirow{2}{*}{$(Z_2,Z_3)$} & A1A$_5$ & 
$(\sin^2 \theta^{\circ}_{13},\sin^2 \theta^{\circ}_{23}) = (0.226,0.436)$ & 
$0.727$ & $\sin^2\th_{23} = 0.554$ \\ 
& A2A$_5$ & $(\sin^2 \theta^{\circ}_{12},\sin^2 \theta^{\circ}_{23}) = (0.226,0.436)$ & 
$-0.727$ & $\sin^2\th_{23} = 0.446$ \\ 
\midrule
$(Z_3,Z_2)$ & B1 & $(\sin^2 \theta^{\circ}_{12},\sin^2 \theta^{\circ}_{23}) = (1/3,1/2)$ 
& $-0.353$ & $\sin^2\th_{12} = 0.341$ \\
\midrule 
$(Z_5,Z_2)$ & B1A$_5$ & $(\sin^2 \theta^{\circ}_{12},\sin^2 \theta^{\circ}_{23}) = (0.276,1/2)$ & 
$-0.405$ & $\sin^2 \theta_{12} = 0.283$ \\ 
\midrule
\multirow{2}{*}{$(Z_2 \times Z_2,Z_2)$} & 
B2A$_5$ & $(\sin^2 \theta^{\circ}_{12},\sin^2 \theta^{\circ}_{13}) = (0.095,0.276)$ & 
$-0.936$ & $\sin^2\th_{12} = 0.331$ \\ 
& B2A$_5$II & $(\sin^2 \theta^{\circ}_{12},\sin^2 \theta^{\circ}_{13}) = (1/4,0.127)$ & 
$1^*$ & $\sin^2\th_{12} = 0.331$  \\ 
\midrule
\multirow{7}{*}{$(Z_2,Z_2)$} & C1 & $\sin^2 \theta^{\circ}_{23} = 1/4$ & $-1^*$ & not fixed\\      
& C3A$_5$ & $\sin^2 \theta^{\circ}_{13} = 0.095$ & $1^*$ & not fixed\\
& C3 & $\sin^2 \theta^{\circ}_{13} = 1/4$ & $-1^*$ & not fixed\\   
& C4A$_5$ & $\sin^2 \theta^{\circ}_{12} = 0.095$ & $-0.799$ & not fixed \\
& C4 & $\sin^2 \theta^{\circ}_{12} = 1/4$ & $1^*$ & not fixed \\
& C8 & $\sin^2 \theta^{\circ}_{23} = 3/4$ & $1^*$ & not fixed \\
& C9A$_5$ & $\sin^2 \theta^{\circ}_{12} = 0.345$ & not fixed & $\sin^2 \theta_{12} = 0.331$ \\
\bottomrule
\end{tabular}
\end{adjustbox}
\caption{The same as in Table~\ref{tab:S4}, but for $G_f = A_5$.}
\label{tab:A5}
\end{table}
%
The exact algebraic forms of the 
irrational values of $\sin^2 \theta^{\circ}_{ij}$ in Table~\ref{tab:A5}
have been found in \cite{Girardi:2015rwa}. 
They are related to the golden ratio $r=(1+\sqrt{5})/2$ 
as follows:  
$2/(4 r^2 - r) \approx 0.226$, 
$r/(6r - 6) \approx 0.436$,
$1/(2 + r) \approx 0.276$,
$1/(4 r^2) \approx 0.095$, 
$1/(3 + 3 r) \approx 0.127$, 
and $(3 - r)/4 \approx 0.345$. 

 We note that case B1 is common to all the three 
flavour symmetry groups $A_4$, $S_4$ and $A_5$, 
while cases C1, C3, C4 and C8 are shared by $S_4$ and $A_5$.
Thus, we have 16 cases in total, which lead to different predictions for 
$\sin^2\th_{12}$ or $\sin^2\th_{23}$ and/or $\cos\delta$. 
As we will see in the next subsection performing a statistical
analysis of these predictions, two cases, namely, C4 and B2A$_5$II, 
are globally disfavoured at more than $3\sigma$ confidence level.
Thus, the total number of phenomenologically viable cases reduces to 14.

\subsection{Statistical Analysis: Current Data}
\label{subsec:present}

 It is important to perform a statistical analysis 
of the predictions for the mixing parameters discussed in the previous subsection 
in order to have a clear picture of their compatibility with the current global 
neutrino oscillation data as well as to
assess the prospects for their future tests.
To this aim, we will follow the method of constructing 
an approximate global likelihood function, which was successfully applied in 
\cite{Marzocca:2013cr,Girardi:2014faa,Girardi:2015vha}
(see also \cite{Girardi:2015zva}).
We briefly describe this method below.

 The NuFIT collaboration performing a global analysis of neutrino oscillation data
provides one-dimensional $\chi^2$ projections for $\sin^2\th_{ij}$ and $\delta$ \cite{NuFITv32Jan2018}. We denote them as $\chi_i^2(x_i)$, $i=1,2,3,4$, 
where $x_i$ are components of 
$\vec{x} = (\sin^2 \theta_{12},\sin^2 \theta_{13},\sin^2 \theta_{23},\delta)$. 
Using these projections, we construct an approximate global $\chi^2$ function as 
\be
\chi^2(\vec{x}) = \sum_{i=1}^4 \chi_i^2(x_i)\,.
\ee
%
For each model (B1, B2S$_4$, C1, etc.), the ``standard'' mixing parameters 
composing vector $\vec{x}$ are not independent, 
but are related to each other via sum rules. 
Thus, in order to obtain a one-dimensional $\chi^2$ function for 
the observable $\alpha$ of interest
($\alpha = \sin^2\th_{12}$, $\sin^2\th_{23}$ or $\cos\delta$), we need to minimise 
the global $\chi^2(\vec{x})$ for each value of $\alpha$ taking into account 
the correlations between the mixing parameters $x_i$ (the sum rules), i.e., 
\be
\chi^2(\alpha) = \min \Bigg[\chi^2(\vec{x}) 
\bigg|_{\stackon[2pt]{\small $\alpha = \mathrm{const}$}
{\small sum rules}}
\Bigg]\,.
\label{eq:chi2alpha}
\ee
%
Finally, we define the global likelihood function as
\be
L(\alpha) = \exp\left(-\frac{\chi^2(\alpha)}{2}\right)\,.
\label{eq:likelihood}
\ee

\begin{figure}
\centering
\includegraphics[width=\textwidth]{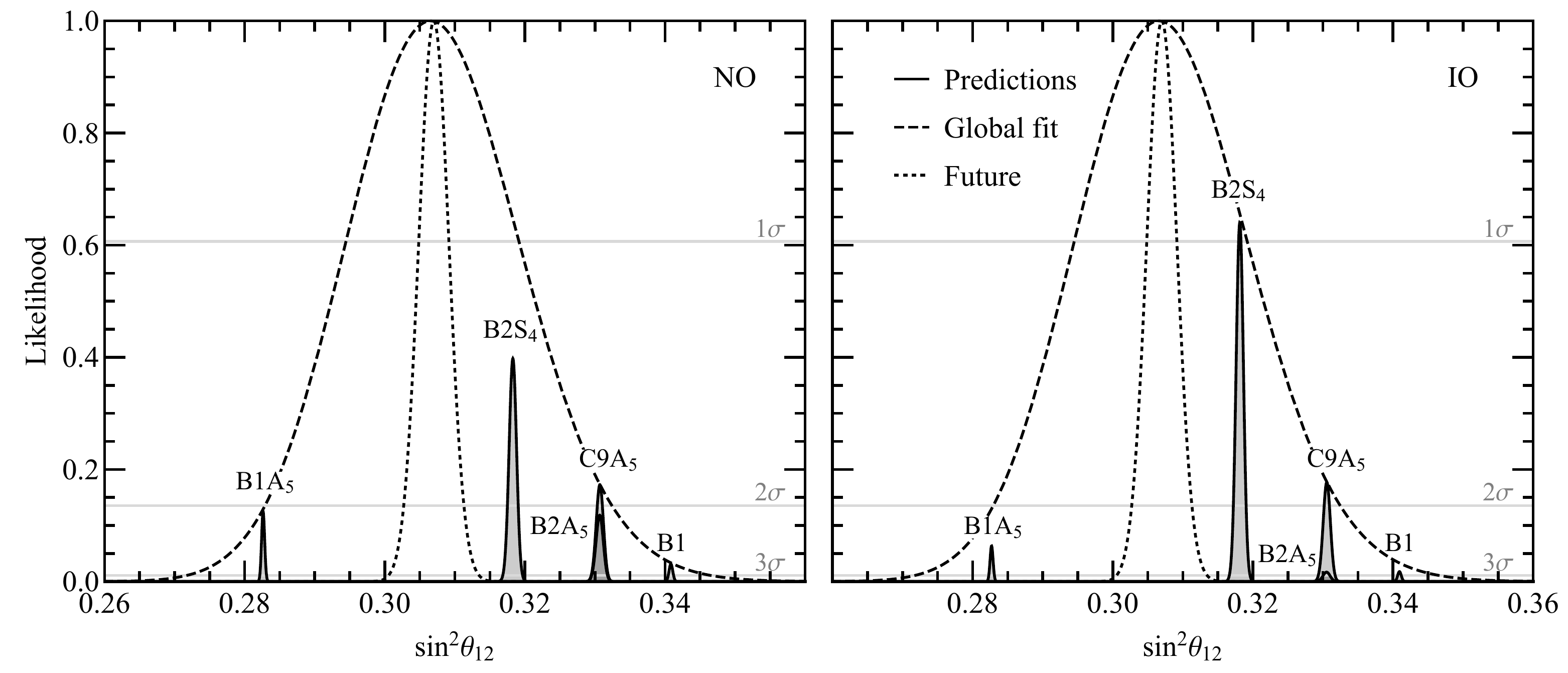}
\caption{Predictions for $\sin^2\th_{12}$ obtained using the current 
global data on the neutrino mixing parameters. 
``Future'' (the dotted line) refers to 
the scenario with $\sin^2\th_{12}^\mathrm{bf} = 0.307$ 
(current best fit value) and the relative $1\sigma$ uncertainty of $0.7\%$ expected from the JUNO experiment. See text for further details.}
\label{fig:ss12present}
\end{figure}

 \textit{Cases predicting $\sin^2\th_{12}$.} 
As can be seen from Tables~\ref{tab:A4}--\ref{tab:A5}, 
there are six different cases which lead to predictions for $\sin^2\th_{12}$. 
Namely, they read B1, B2S$_4$, B1A$_5$, B2A$_5$, B2A$_5$II and C9A$_5$. 
We have performed the statistical analysis of the predictions for $\sin^2\th_{12}$ 
as described above. In Fig.~\ref{fig:ss12present}, we present 
the obtained likelihood functions. 
%
In the left (right) panel, we have used as an input the one-dimensional projections 
$\chi_i^2(x_i)$ for NO (IO). 
We would like to note that according to \cite{NuFITv32Jan2018} 
there is an overall preference for NO over IO of $\D\chi^2 = 4.14$. 
However, we take a conservative approach and treat both orderings 
on equal grounds in our analysis. 

 Five cases presented in Fig.~\ref{fig:ss12present} 
lead to very sharp predictions for $\sin^2\th_{12}$. 
The corresponding likelihood profiles are very narrow  
because their widths are determined by the small uncertainty on $\sin^2\th_{13}$ 
as can be understood from eqs.~\eqref{eq:ss12B1}, \eqref{eq:ss12B2} and \eqref{eq:ss12C9}. 
Case B1 is compatible with the global data at $3\sigma$. 
Cases B1A$_5$ and B2A$_5$ almost touch the $2\sigma$ line for NO 
and are within $3\sigma$ for IO. 
C9A$_5$ is compatible with the data at $2\sigma$.  
Finally, B2S$_4$ is the case which is favoured most by the present data, 
being compatible with them at $1.5\sigma$ for NO and $1\sigma$ for IO.
We find that case B2A$_5$II is globally disfavoured at more than $3\sigma$, 
the value of $\chi^2$ in the minimum, $\chi^2_\mathrm{min}$, 
being equal to 9.9 (13.7) for NO (IO). 
Thus, we do not present this case in Fig.~\ref{fig:ss12present}. 

 The dashed line corresponds to the likelihood for $\sin^2\th_{12}$ 
extracted from the global analysis, i.e., calculated substituting 
the one-dimensional projection $\chi_1^2(\sin^2\th_{12})$ in eq.~\eqref{eq:likelihood}
in place of $\chi^2(\alpha)$. 
It is clear from the way in which the likelihood function is constructed 
that none of the predicted likelihood profiles can go beyond the dashed line. 
The dotted line instead represents 
the prospective precision on $\sin^2\th_{12}$ of $0.7\%$, 
which is planned to be achieved by 
the medium-baseline reactor oscillation experiment JUNO \cite{An:2015jdp}. 
More precisely, the corresponding likelihood is calculated using eq.~\eqref{eq:likelihood} 
with a replacement of $\chi^2(\alpha)$ by 
\be
\chi_\mathrm{1,\,future}^2\left(\sin^2\th_{12}\right) = \left(
\frac{\sin^2\th_{12} - \sin^2\th_{12}^\mathrm{bf}}{\sigma(\sin^2\th_{12})}
\right)^2\,,
\label{eq:chi2ss12future}
\ee
%
where $\sin^2\th_{12}^\mathrm{bf} = 0.307$ is the current best fit value of $\sin^2\th_{12}$, 
and $\sigma(\sin^2\th_{12}) = 0.007 \times \sin^2\th_{12}^\mathrm{bf}$ 
is the prospective $1\sigma$ uncertainty in its determination.
Thus, we make an assumption that the best fit value of $\sin^2\th_{12}$ 
will not change in the future. 
If it is indeed the case, then, as is clear from Fig.~\ref{fig:ss12present}, 
all five models, B1, B2S$_4$, B1A$_5$, B2A$_5$ and C9A$_5$, 
will be ruled out by the JUNO measurement of $\sin^2\th_{12}$. 
If, however, the best fit value changed coinciding, e.g., with that of case B1A$_5$ (B2S$_4$), 
cases B2S$_4$ (B1A$_5$), B2A$_5$, C9A$_5$ and B1 would be ruled out.

\begin{figure}
\centering
\includegraphics[width=\textwidth]{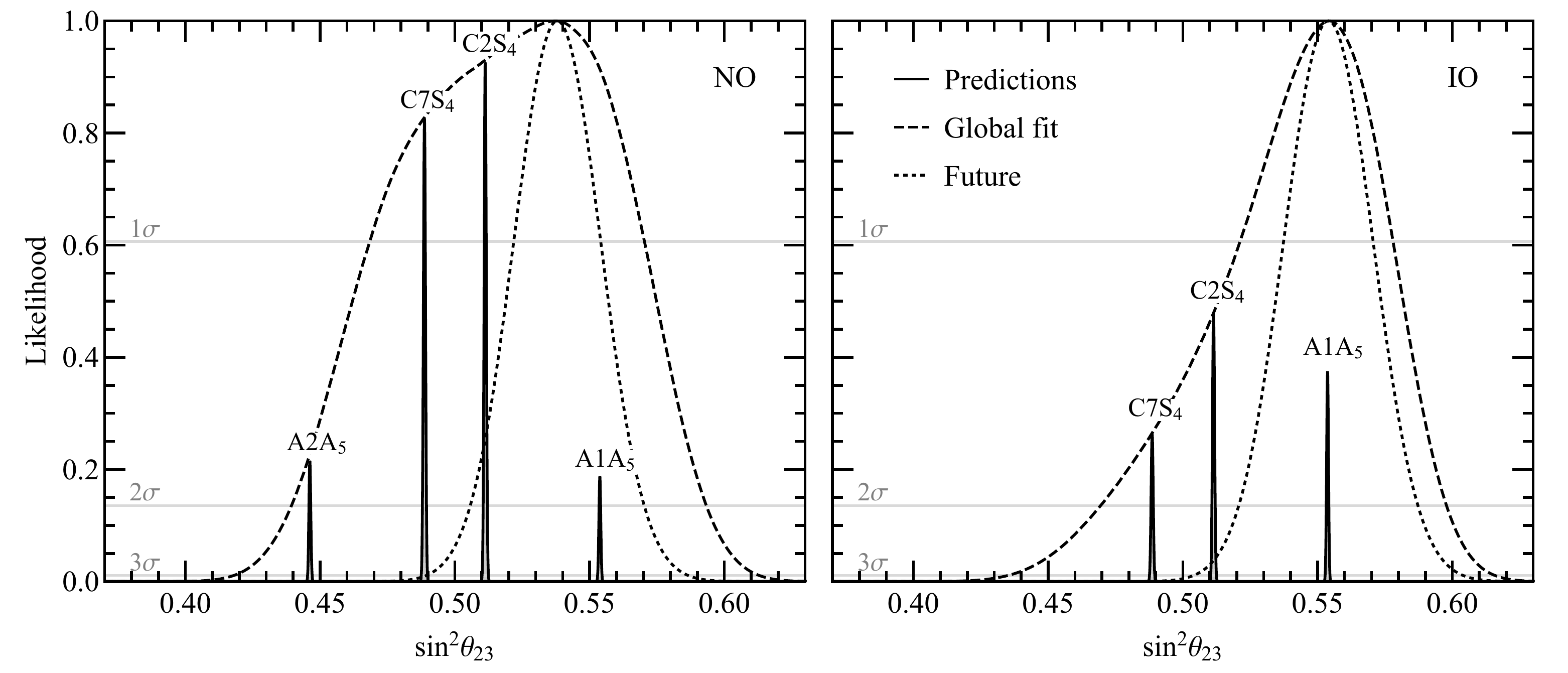}
\caption{Predictions for $\sin^2\th_{23}$ obtained using the current 
global data on the neutrino mixing parameters. 
``Future'' (the dotted line) refers to the scenario with 
$\sin^2\th_{23}^\mathrm{bf} = 0.538~(0.554)$ 
for NO (IO) (current best fit values) and the relative $1\sigma$ uncertainty of $3\%$ expected from DUNE and T2HK. See text for further details.}
\label{fig:ss23present}
\end{figure}

 \textit{Cases predicting $\sin^2\th_{23}$.} 
There are four cases leading to predictions for $\sin^2\th_{23}$: 
C2S$_4$, C7S$_4$, A1A$_5$ and A2A$_5$. 
We show the corresponding likelihood functions in Fig.~\ref{fig:ss23present}. 
%
Since, in these cases $\sin^2\th_{23}$ is determined by $\sin^2\th_{13}$, see 
eqs.~\eqref{eq:ss23A1}, \eqref{eq:ss23A2}, \eqref{eq:ss23C2} and \eqref{eq:ss23C7}, 
the predicted likelihood profiles are very narrow. 
Cases C2S$_4$ and C7S$_4$ are well compatible with the data for NO 
(at less than $1\sigma$) and with the data for IO (at around $1.5\sigma$). 
What concerns cases A1A$_5$ and A2A$_5$, 
they reconcile with the data for NO at $2\sigma$. 
For IO, A1A$_5$ is within $1.5\sigma$, while A2A$_5$ is disfavoured at more than $3\sigma$ ($\chi^2_\mathrm{min} = 10.1$).
This is why this case is not present in the right panel of Fig.~\ref{fig:ss23present}.

 Similarly to the previous figure, the dashed line corresponds to the global fit likelihood 
obtained from the one-dimensional projection $\chi_3^2(\sin^2\th_{23})$. 
The dotted line indicates the prospective precision on $\sin^2\th_{23}$ of $3\%$.
It is worth noting that the error on $\sin^2\th_{23}$, which can be reached 
in the next generation of long-baseline (LBL) neutrino oscillation experiments 
like DUNE \cite{Acciarri:2016crz,Acciarri:2015uup} 
and T2HK \cite{Abe:2014oxa,Abe:2015zbg}, 
depends on the true value of this parameter. As can be seen, e.g., 
from Fig.~10 in \cite{Ballett:2016daj}, 
in the case of T2HK this error varies from $1\%$ 
for the true values of $\sin^2\th_{23}$ on the boundaries of its $3\sigma$ range 
to approximately $6\%$ for $\sin^2\th_{23} = 0.5$. 
For the current best fit value of $\sin^2\th_{23} = 0.538$ (for NO), 
the expected uncertainty does not exceed $3\%$, 
and we take it as a benchmark value. 
The likelihood corresponding to the dotted line is calculated using  
\be
\chi_\mathrm{3,\,future}^2\left(\sin^2\th_{23}\right) = \left(
\frac{\sin^2\th_{23} - \sin^2\th_{23}^\mathrm{bf}}{\sigma(\sin^2\th_{23})}
\right)^2\,,
\label{eq:chi2ss23future}
\ee
%
where $\sin^2\th_{23}^\mathrm{bf} = 0.538~(0.554)$
is the current best fit value of $\sin^2\th_{23}$ for NO (IO), 
and $\sigma(\sin^2\th_{23}) = 0.03 \times \sin^2\th_{23}^\mathrm{bf}$ 
is the prospective $1\sigma$ uncertainty.
If the current best fit value does not change in the future, 
case A2A$_5$ will be ruled out, 
while case C7S$_4$ will be disfavoured at $3\sigma$. 
However, if the best fit value changed, e.g., to 0.5 for both NO and IO spectra, 
cases C2S$_4$ and C7S$_4$ would be phenomenologically viable, 
while cases A1A$_5$ and A2A$_5$ would be disfavoured at $3\sigma$ 
(see Fig.~\ref{fig:ss23present}).

\begin{figure}
\centering
\includegraphics[width=\textwidth]{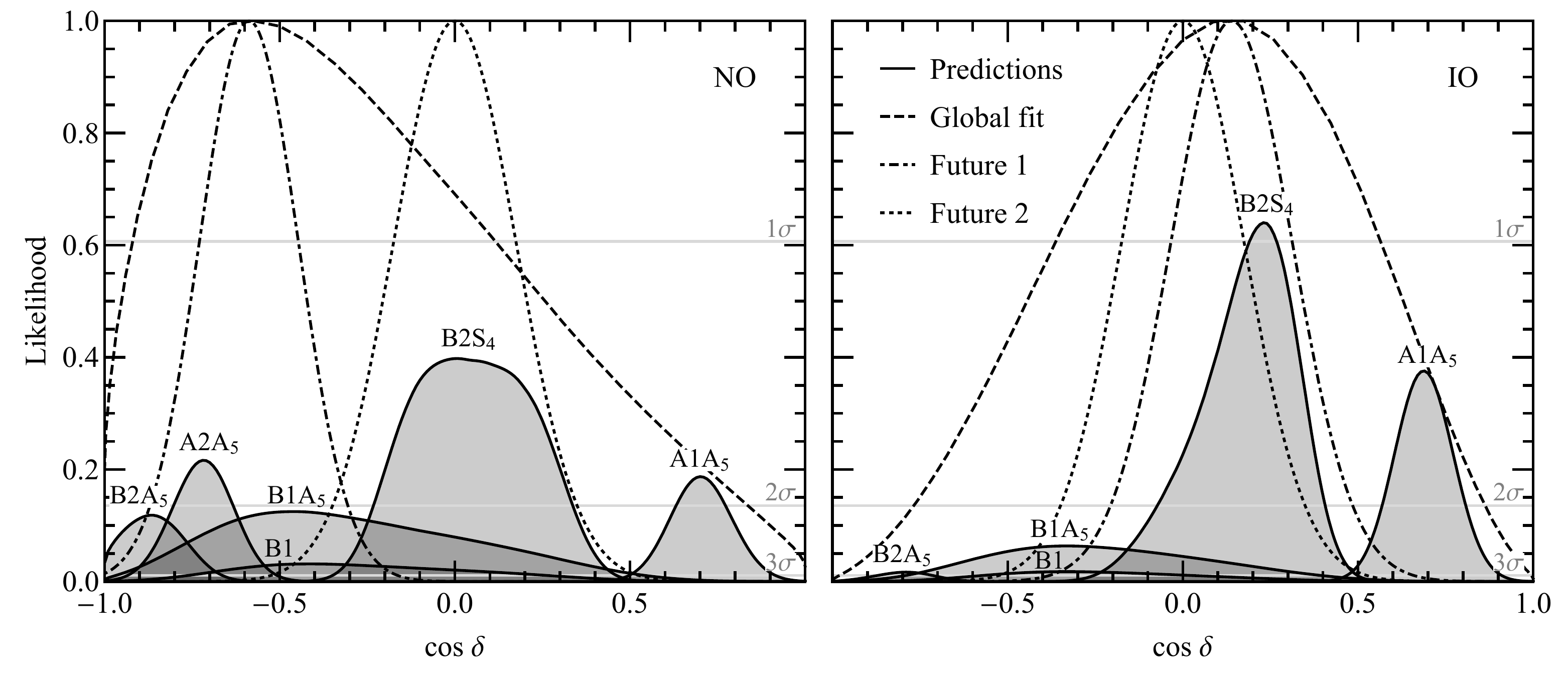}
\caption{Predictions for $\cos\delta$ in viable cases A and B 
obtained using the current global data on 
the neutrino mixing parameters. 
``Future 1'' (the dash-dotted line) refers to the scenario with $\delta^\mathrm{bf} = 234\degree~(278\degree)$ for NO (IO) (current best fit values) 
and the $1\sigma$ uncertainty on $\delta$ of $10\degree$. 
``Future 2'' (the dotted line) corresponds to $\delta^\mathrm{bf} = 270\degree$ and 
the $1\sigma$ uncertainty on $\delta$ of $10\degree$. 
See text for further details.}
\label{fig:cosdeltaABpresent}
\end{figure}

 \textit{Cases predicting $\cos\delta$.} 
As has been discussed in Section~\ref{sec:patterns} 
and can be seen from Tables~\ref{tab:A4}--\ref{tab:A5}, 
cases A and B of interest lead not only to predictions for 
$\sin^2\th_{23}$ and $\sin^2\th_{12}$, respectively, but also to 
predictions for $\cos\delta$. 
Using eqs.~\eqref{eq:ss23A1}--\eqref{eq:cosdeltaB2}, we have performed 
the statistical analysis of these predictions. 
The obtained results are summarised in Fig.~\ref{fig:cosdeltaABpresent}.
%
We find that the predictions for $\cos\delta$ in cases B are very sensitive to 
the value of $\th_{23}$ (cf. eqs.~\eqref{eq:cosdeltaB1} 
and \eqref{eq:cosdeltaB2}), which is 
determined with a larger uncertainty than $\th_{12}$ and $\th_{13}$.
This results in quite broad likelihood profiles.
For cases A, the uncertainty in predicting $\cos\delta$ 
from eqs.~\eqref{eq:cosdeltaA1} and \eqref{eq:cosdeltaA2} is driven by 
the uncertainty on $\sin^2\th_{12}$,  
since $\sin^2\th_{23}$ is almost fixed in these cases 
(see Fig.~\ref{fig:ss23present}). 
Thus, the resulting likelihood profiles are not so broad in cases 
A1A$_5$ and A2A$_5$.
In each case B (A), the value of the likelihood in the maximum is the same as 
in Fig.~\ref{fig:ss12present} (Fig.~\ref{fig:ss23present}) as should 
be expected from the procedure of constructing the likelihood.

 The dashed line in Fig.~\ref{fig:cosdeltaABpresent} stands for the likelihood 
extracted from the global analysis. More precisely, we take the 
one-dimensional projection 
$\chi_4^2(\delta)$ restricted to the interval 
of $\delta\in[180\degree,360\degree]$ 
and translate it to $\chi_4^2(\cos\delta)$. Then, we use the latter 
to construct the likelihood.
At present, all values of $\cos\delta$ are allowed at $3\sigma$ for NO, 
and almost all, $\cos\delta\in[-0.978,0.995]$, for IO. 
We also show the dash-dotted and dotted lines which represent two benchmark cases. 
The first case, marked in Fig.~\ref{fig:cosdeltaABpresent} as ``Future 1'' 
(the dash-dotted line), corresponds to  
the current best fit value $\delta^\mathrm{bf} = 234\degree~(278\degree)$ for NO (IO) 
and the prospective $1\sigma$ uncertainty $\sigma(\delta) = 10\degree$.
The second case, ``Future 2'' (the dotted line), corresponds to 
the potential best fit value $\delta^\mathrm{bf} = 270\degree$ 
(for both NO and IO) 
and the same error on $\delta$ of $10\degree$. 
The corresponding $\chi^2$ functions read 
\be
\chi_\mathrm{4,\,future}^2\left(\cos\delta\right) = \left(
\frac{\cos\delta - \cos\delta^\mathrm{bf}}{\sigma(\cos\delta)}
\right)^2\,,
\ee
%
where $\sigma(\cos\delta)$ is obtained from $\sigma(\delta) = 10\degree$ 
using the derivative method of uncertainty propagation.

\begin{figure}
\centering
\includegraphics[width=\textwidth]{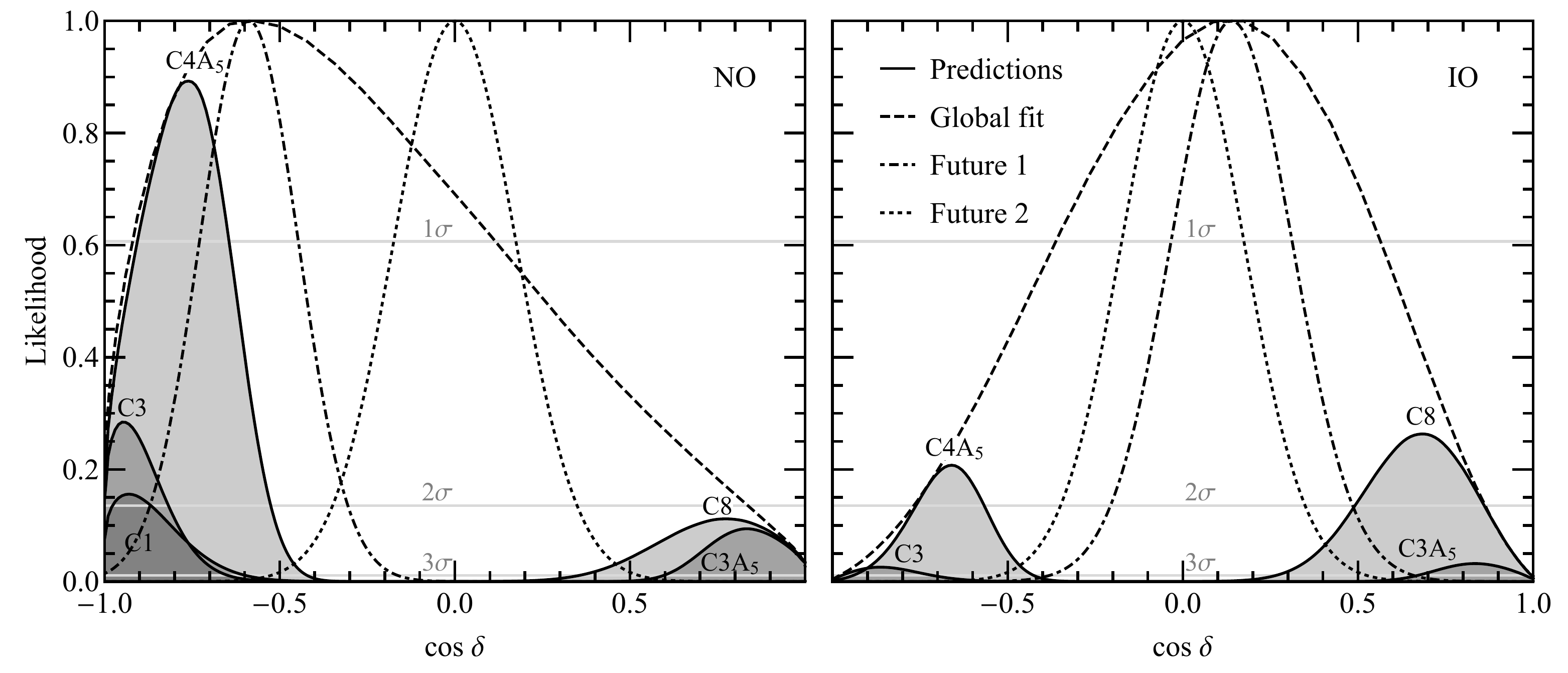}
\caption{The same as in Fig.~\ref{fig:cosdeltaABpresent}, but for viable cases C.}
\label{fig:cosdeltaCpresent}
\end{figure}

 Finally, we perform the statistical analysis of the predictions 
for $\cos\delta$ 
in cases C1, C3, C4, C8, C3A$_5$ and C4A$_5$. The corresponding sum rules are 
given in eqs.~\eqref{eq:cosdeltaC1}--\eqref{eq:cosdeltaC8}. 
Note that none of the mixing angles are predicted in these cases. 
We show the obtained likelihood functions for 
$\cos\delta$ in Fig.~\ref{fig:cosdeltaCpresent}.
%
As we see, all of them peak at values of $|\cos\delta|\sim0.5-1$. 
There are two groups of cases: 
the first one consisting of C1, C3 and C4A$_5$ leads to the negative values of $\cos\delta$, 
while the second one including C8 and C3A$_5$ predicts the positive values. 
We find that case C4 is globally disfavoured at more than $3\sigma$, 
the corresponding $\chi^2_\mathrm{min}$ being $9.3$ ($13.6$) for NO (IO). 
Therefore, we do not present this case in Fig.~\ref{fig:cosdeltaCpresent}.
On contrary, case C4A$_5$ is very well compatible with the data for NO, 
while for IO the compatibility is somewhat worse, at around $2\sigma$. 
Case C3 reconciles with the data for NO (IO) 
at approximately $1.5\sigma$ ($3\sigma$).  
Case C1, being compatible at $2\sigma$ for NO, 
gets disfavoured at more than $3\sigma$ for IO, 
the corresponding $\chi^2_\mathrm{min} = 12.7$. 
C8 is concordant with the data at almost $2\sigma$ ($1.5\sigma$) for NO (IO).
Finally, the predictions of C3A$_5$ are compatible with 
the global data at $3\sigma$. 

 Looking at the dotted line, we see that if in the future 
the best fit value of $\delta$ 
shifted to $270\degree$ and the LBL experiments managed to achieve 
the $1\sigma$ uncertainty on $\delta$ of $10\degree$, 
cases C1, C3 and C3A$_5$ (C4A$_5$ and C8) would 
be disfavoured at more than (at around) 
$3\sigma$ only by the measurement of $\delta$.
If, however, the current best fit value of $\delta$ for the NO spectrum 
is shown to be the true value for both the NO and IO spectra,
cases  C3A$_5$ and C8 will be ruled out 
by the measurement of $\delta$ with the indicated precision.
In addition, the precision on $\sin^2\th_{12}$ and $\sin^2\th_{23}$ 
will be also improved. 
This will modify the likelihood profiles making them narrower.
In the next subsection, we will study how this improvement will affect the results 
presented in Figs.~\ref{fig:ss12present}--\ref{fig:cosdeltaCpresent}.

\subsection{Statistical Analysis: Prospective Data}
\label{subsec:future}

 In this subsection, we want to access the impact of the 
future precision measurements 
of the neutrino mixing angles on the predictions discussed 
in subsection~\ref{subsec:present}.
To this aim, we perform a statistical analysis of these 
predictions assuming that 
(i) the current best fit values of the mixing angles 
will not change in the future, 
and (ii) the prospective relative $1\sigma$ uncertainties on 
$\sin^2\th_{12}$, $\sin^2\th_{23}$ and $\sin^2\th_{13}$ 
will amount to $0.7\%$, $3\%$ and $3\%$, respectively.
As has already been mentioned, a measurement of $\sin^2\th_{12}$ 
with such a high precision is expected from JUNO, while DUNE and T2HK 
will be able to reach $3\%$ on $\sin^2\th_{23}$ if atmospheric mixing 
deviates somewhat from maximal 
(see the discussion above eq.~\eqref{eq:chi2ss23future}).
What concerns the reactor angle, 
Daya Bay is going to attain the precision of $3\%$ on $\sin^2\th_{13}$ 
by the year of 2020 \cite{Ling:2016wgq}.
The results of the analysis in this subsection should be considered only as indicative. 
Similar analysis should be performed when real data become available.
 
 With these assumptions, we construct a global 
$\chi^2_\mathrm{future}$ function as
\be
\chi^2\left(\vec{y}\right) = \sum_{i=1}^3 \chi_{i,\,\mathrm{future}}^2\left(y_i\right)\,,
\ee
%
where $\vec{y} = (\sin^2\th_{12},\sin^2\th_{13},\sin^2\th_{23})$, 
the functions $\chi_{i,\,\mathrm{future}}^2(y_i)$ with $i=1$ and $i=3$ 
are given in eqs.~\eqref{eq:chi2ss12future} and \eqref{eq:chi2ss23future}, 
respectively, and we define $\chi_{2,\,\mathrm{future}}^2(\sin^2\th_{13})$ as
\be
\chi_\mathrm{2,\,future}^2\left(\sin^2\th_{13}\right) = \left(
\frac{\sin^2\th_{13} - \sin^2\th_{13}^\mathrm{bf}}{\sigma(\sin^2\th_{13})}
\right)^2\,,
\label{eq:chi2ss13future}
\ee
%
with $\sin^2\th_{13}^\mathrm{bf} = 0.02206~(0.02227)$ being the current 
best fit value of $\sin^2\th_{13}$ for NO (IO), 
and $\sigma(\sin^2\th_{13}) = 0.03\times\sin^2\th_{13}^\mathrm{bf}$ being 
the prospective $1\sigma$ uncertainty in its determination.
We note that by constructing $\chi^2_\mathrm{future}$ in this way, 
we do not assume any experimental input on $\delta$. 
We use $\chi^2_\mathrm{future}(\vec{y})$ instead of $\chi^2(\vec{x})$ 
in eq.~\eqref{eq:chi2alpha} to construct $\chi^2(\alpha)$. 
Finally, the likelihood function is calculated according 
to eq.~\eqref{eq:likelihood}.

 \textit{Cases predicting $\sin^2\th_{12}$.} 
As we have already mentioned earlier, it is clear from 
Fig.~\ref{fig:ss12present} that
JUNO will be able to rule out all the cases predicting $\sin^2\th_{12}$, 
if the best fit value of this parameter does not shift in 
the future (see the dotted line). 
However, this conclusion might change if the best fit value of $\sin^2\th_{12}$ 
changes significantly.

\begin{figure}
\centering
\includegraphics[width=\textwidth]{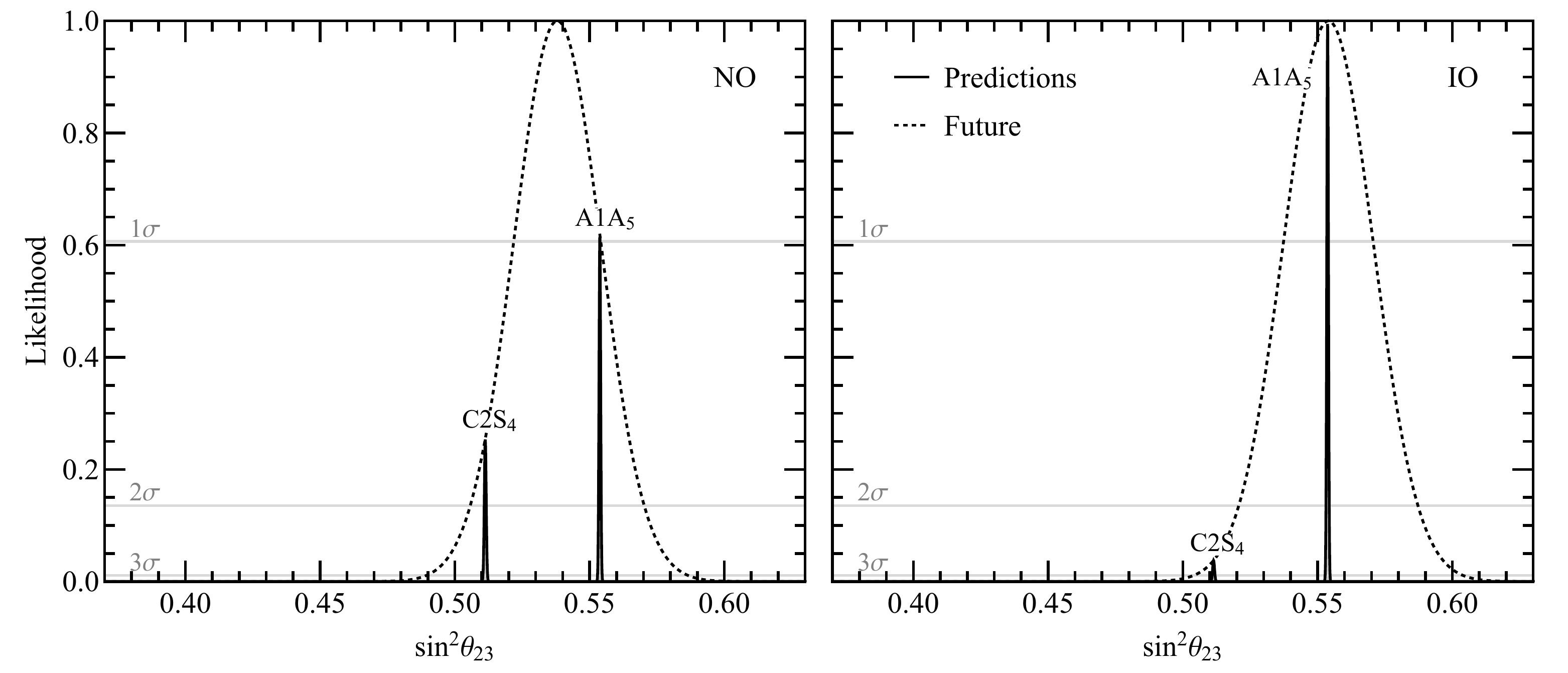}
\caption{Predictions for $\sin^2\th_{23}$ obtained using 
the current best fit values and the prospective uncertainties in the determination 
of the neutrino mixing angles. 
``Future'' (the dotted line) refers to the scenario with $\sin^2\th_{12}^\mathrm{bf} = 0.307$ 
(current best fit value) and the relative $1\sigma$ uncertainty of $0.7\%$ expected from the JUNO experiment. See text for further details.}
\label{fig:ss23future}
\end{figure}

 \textit{Cases predicting $\sin^2\th_{23}$.} 
Since the predicted centre value of $\sin^2\th_{23} = 0.554$ in case A1A$_5$ 
matches exactly the current best fit value of this parameter for IO, 
this case will certainly survive in the future, 
if $\sin^2\th_{23}^\mathrm{bf}$ remains the same.
Moreover, the precision on $\sin^2\th_{23}$ is not expected to be as high as on $\sin^2\th_{12}$, and we can infer from Fig.~\ref{fig:ss23present} that 
case C2S$_4$ has a chance to survive, while A2A$_5$ and C7S$_4$ do not. 
We have performed the statistical analysis with the prospective uncertainties. 
The obtained results presented in Fig.~\ref{fig:ss23future} confirm our expectations.
%
In particular, case A1A$_5$ would be perfectly compatible with the 
prospective data for IO. Note that now the amplitude of the likelihood profile 
is maximal, since we have not assumed any information on $\delta$. 
For NO, the case under consideration would be slightly disfavoured
only due to the form of $\chi_\mathrm{3,\,future}^2\left(\sin^2\th_{23}\right)$ 
(the dotted line). 
C2S$_4$ would be compatible at $2\sigma$ ($3\sigma$) with the 
prospective data for NO (IO), which is again dictated by the dotted line. 
For C7S$_4$ we find $\chi^2_\mathrm{min} = 9.3~(15.5)$ for NO (IO), 
and thus, we do not present this case in Fig.~\ref{fig:ss23future}.
The conclusions about the excluded cases should be revised if 
the best fit value of $\sin^2\th_{23}$ shifts, e.g., to 0.5.

\begin{figure}
\centering
\includegraphics[width=\textwidth]{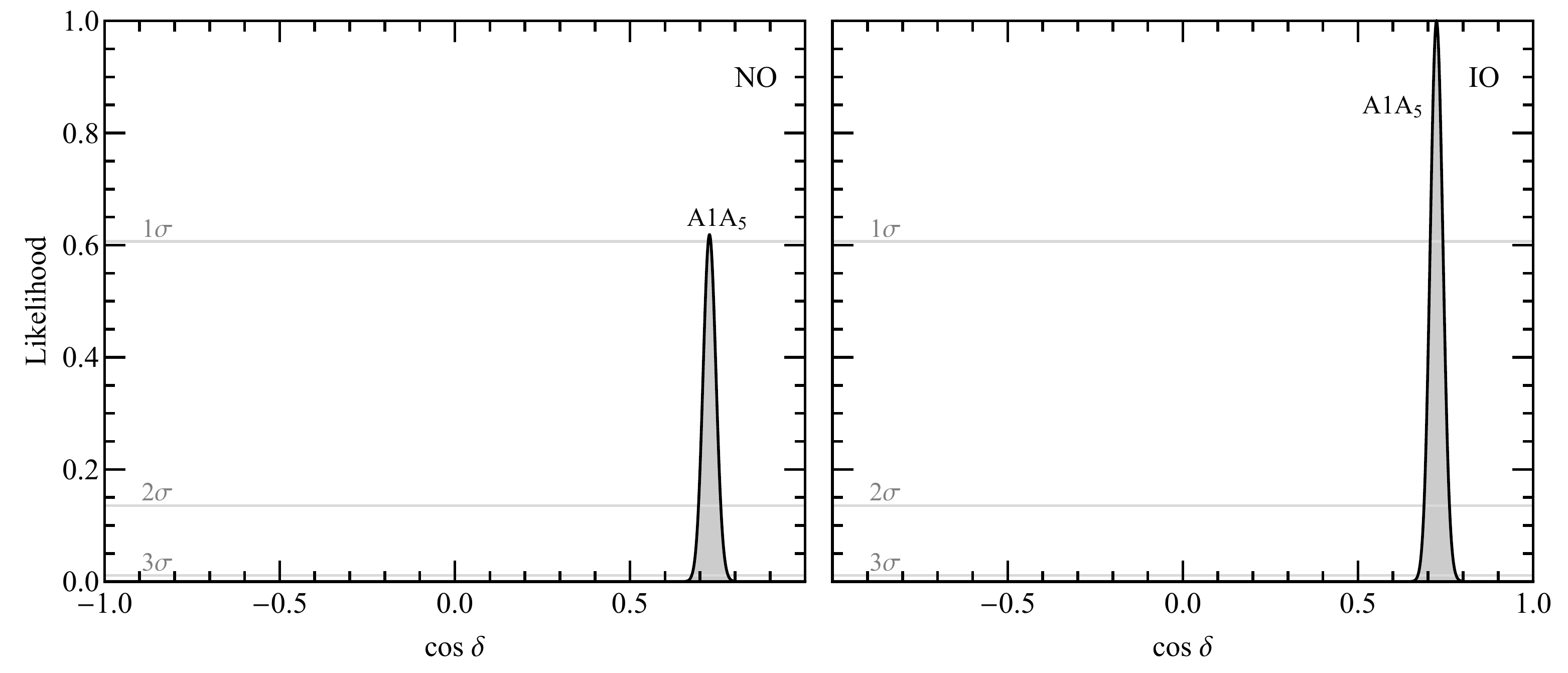}
\caption{Predictions for $\cos\delta$ in only viable case A1A$_5$
obtained using 
the current best fit values and the prospective uncertainties in the determination 
of the neutrino mixing angles. 
See text for further details.}
\label{fig:cosdeltaABfuture}
\end{figure}

 \textit{Cases predicting $\cos\delta$.} 
Since all cases B as well as case A2A$_5$ would be ruled out by the prospective 
data we have assumed, 
Fig.~\ref{fig:cosdeltaABpresent} would change significantly in the future, 
featuring \textit{only} case A1A$_5$. 
We present the likelihoods obtained in this case for NO and IO 
in Fig.~\ref{fig:cosdeltaABfuture}.
%
The width of the likelihood profiles in this figure is much smaller than 
that of the corresponding profiles in Fig.~\ref{fig:cosdeltaABpresent}. 
This makes even more evident the fact that improving the precision on 
the mixing angles leads to sharper predictions for $\cos\delta$,  
which can and should be considered as an additional motivation of 
measuring the mixing angles with a high precision.

\begin{figure}[t!]
\centering
\includegraphics[width=\textwidth]{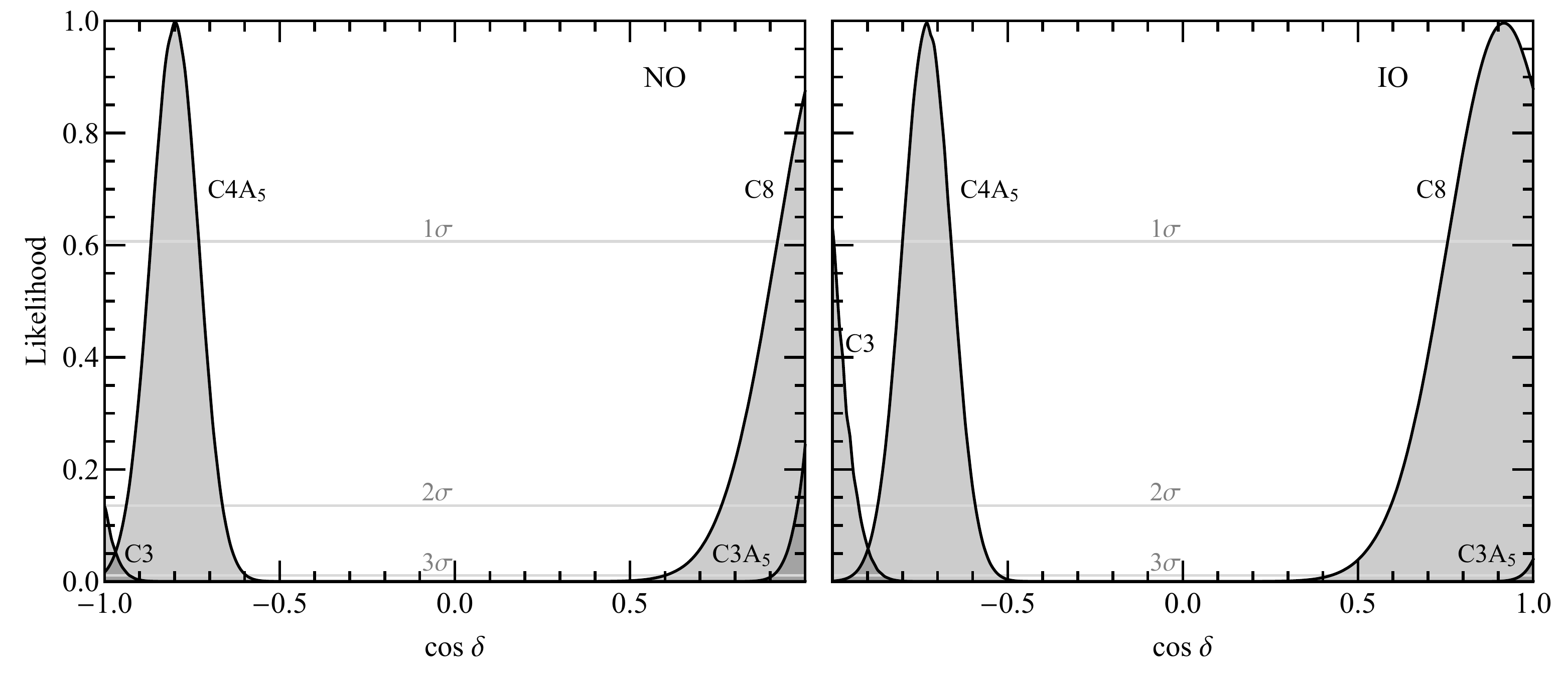}
\caption{The same as in Fig.~\ref{fig:cosdeltaABfuture}, but for viable cases C.}
\label{fig:cosdeltaCfuture}
\end{figure}

 Finally, we perform the statistical analysis of the predictions 
for $\cos\delta$ in cases C. 
We show the results in Fig.~\ref{fig:cosdeltaCfuture}.
%
We find that under the assumptions made case C1 would be ruled out.
Thus, we would be left with four cases.
Two of them lead to predictions 
which are in the corners of the parameter space for $\cos\delta$. 
Namely, C3 leads to values of $\cos\delta \ltap {-0.9}~(-0.8)$ for NO (IO), 
while C3A$_5$ leads to $\cos\delta \gtap 0.9$. 
At least some of these values, if not all of them, 
will be ruled out by the future data on $\delta$.
In what concerns currently viable cases  C4A$_5$ and C8,
they will be disfavoured at approximately $3\sigma$ only by 
the measurement of $\delta$ if the true value of 
$\delta$ is indeed around $270\degree$ 
and the planned LBL experiments measure 
$\delta$ with a $1\sigma$ error of $10\degree$ 
(cf. Fig.~\ref{fig:cosdeltaCpresent}). 
At the same time, if the current best fit value of $\delta$ for 
the NO spectrum turned out to be the true value for both 
the NO and IO spectra, cases C3 and C4A$_5$ would ``survive'' this test.
Thus, a high precision measurement of $\delta$ is crucial to 
firmly establish the status of the considered cases.

 Before concluding, let us add two comments. 
First, the predictions considered in the present study can be tested 
simulating the future neutrino oscillation experiments, as 
it has been recently done, e.g., in ref.~\cite{Agarwalla:2017wct},  
where DUNE and T2HK simulations have been performed to test
the predictions for $\cos\delta$ of sum rules \cite{Petcov:2014laa} 
corresponding to pattern D of discrete flavour symmetry breaking 
(see the Introduction). We plan to present such a study elsewhere. 
Secondly, it has been shown in ref.~\cite{Gehrlein:2016fms} 
for the indicated set of sum rules 
that renormalisation group corrections to their predictions 
are negligible within the SM extended by the Weinberg dimension 5 operator 
to generate the neutrino masses, 
as well as in the MSSM with relatively 
small $\tan\beta$ and the lightest neutrino mass $\ll 0.01$~eV. 
The renormalisation group corrections can be sizeable in 
the MSSM if these conditions are not fulfilled.

\section{Conclusions}
\label{sec:conclusions}

 We have investigated the phenomenological viability of the discrete (lepton) flavour
symmetries $A_4$, $S_4$ and $A_5$ for the description of neutrino mixing. 
More specifically, we have considered the $A_4$, $S_4$ and $A_5$ 
lepton flavour symmetry groups broken to non-trivial residual symmetry
subgroups $G_e$ and $G_\nu$ in the charged lepton and neutrino sectors. 
All flavour symmetry breaking patterns considered by us 
involve a $Z_2$ group as a residual symmetry in one of the 
two sectors, or two different $Z_2$ groups as residual symmetries in both sectors. 
More precisely, these patterns read: 
(A) $G_e = Z_2$ and $G_{\nu} = Z_k$, $k > 2$ or $Z_m \times Z_n$, $m,n \geq 2$; 
(B)~$G_e = Z_k$, $k > 2$ or $Z_m \times Z_n$, $m,n \geq 2$ and $G_{\nu} = Z_2$;
and (C) $G_e = Z_2$ and $G_{\nu} = Z_2$. 
In the cases corresponding to pattern A (B) sum rules for 
$\sin^2\th_{23}$ ($\sin^2\th_{12}$) and $\cos\delta$ arise, 
while pattern C leads to sum rules for either 
$\sin^2\th_{12}$ or  $\sin^2\th_{23}$ or $\cos\delta$ \cite{Girardi:2015rwa}, 
$\theta_{12}$, $\theta_{23}$ and $\delta$ being the solar, 
atmospheric neutrino mixing angles and the Dirac 
CP violation phase of the PMNS neutrino mixing matrix.
 
 We have performed a statistical analysis of the sum rule predictions 
using as input the latest global neutrino oscillation 
data \cite{NuFITv32Jan2018}.
We have found 14 cases in total compatible with these data at 
$3\sigma$ confidence level.
Five of them lead to very sharp predictions for $\sin^2\th_{12}$, 
and four others to similarly sharp 
predictions for $\sin^2\th_{23}$ 
(see Figs.~\ref{fig:ss12present} and \ref{fig:ss23present}). 
Phenomenologically viable
cases A and B, which are six in total, 
lead as well to predictions for $\cos\delta$ 
presented in Fig.~\ref{fig:cosdeltaABpresent}.
Five viable C cases also lead to predictions 
for $\cos\delta$, which are summarised  
in Fig.~\ref{fig:cosdeltaCpresent}.
The corresponding likelihoods 
peak at values of $|\cos\delta| \sim 0.5-1$. 
As we have shown, the number of these cases could be further 
reduced by a sufficiently precise measurement of $\delta$.

 Further, we have performed a statistical analysis of 
the predictions discussed above assuming that 
(i) the current best fit values of the mixing angles 
will not change in the future, 
and (ii) the prospective relative $1\sigma$ uncertainties on 
$\sin^2\th_{12}$, $\sin^2\th_{23}$ and $\sin^2\th_{13}$ 
will amount to $0.7\%$, $3\%$ and $3\%$, respectively.
Such uncertainties are planned to be achieved by 
the JUNO, T2HK/DUNE and Daya Bay experiments, respectively.
Under the assumptions made, all the cases predicting $\sin^2\th_{12}$ 
(see Fig.~\ref{fig:ss12present}) get ruled out. 
In what concerns the cases predicting $\sin^2\th_{23}$, two out of
the four would ``survive'' this test  
(Fig.~\ref{fig:ss23future}). 
We have found that only one case among six cases A and B  
viable at present 
would be compatible with the prospective data on the neutrino mixing angles. 
The predictions for $\cos\delta$ in this case are shown in Fig.~\ref{fig:cosdeltaABfuture}.
Four out of five cases C predicting $\cos\delta$ 
satisfy the expected constraints on the mixing angles. 
The corresponding predictions are summarised in Fig.~\ref{fig:cosdeltaCfuture}. 
Thus, in total six cases out of 14 viable at present are compatible 
with the assumed prospective data on the neutrino mixing angles,
provided the current best fit values of the three neutrino 
mixing angles will not change drastically in the future. 
Five of these cases will be further 
critically tested by sufficiently precise data on 
the Dirac phase $\delta$, e.g., if $\delta$ is measured with 
$1\sigma$ uncertainty of $10\degree$.
Obviously, the results obtained with the prospective data 
might change with the accumulation of new data 
if, e.g., the current best fit values of $\sin^2\th_{12}$ and/or $\sin^2\th_{23}$ 
change significantly.

 In summary, we have shown that  the $A_4$, $S_4$ and $A_5$ 
lepton flavour symmetries, broken to non-trivial residual symmetries
in the charged lepton and neutrino sectors,
lead in the case of 3-neutrino mixing
to a relatively small number of phenomenologically 
viable cases characterised by distinct predictions for 
the solar or atmospheric neutrino 
mixing angles $\theta_{12}$ and $\theta_{23}$  and/or for 
the cosine of the Dirac CP violation phase $\delta$. 
We have also shown that the high precision measurements 
of the three neutrino 
mixing angles, planned to be performed by Daya Bay 
and the next generation of neutrino 
oscillation experiments~---~JUNO, T2HK, DUNE~---~%
can reduce the number of the phenomenologically 
viable cases to six. Five of these cases will be further 
critically tested by sufficiently precise data on 
the Dirac phase $\delta$ that could be provided by the 
T2HK and DUNE experiments.

 The results obtained in the present study 
show that the future high precision data 
on the three neutrino mixing angles and on 
the leptonic Dirac CP violation phase $\delta$, planned 
to be obtained in the Daya Bay, T2K, NO$\nu$A, 
and especially by the JUNO, T2HK and DUNE experiments,
will be crucial for testing the ideas 
of existence of new fundamental underlying  discrete (non-Abelian) symmetry 
of the PMNS neutrino mixing matrix and of the lepton sector of particle physics.

\section*{Acknowledgements}

 We would like to thank I.~Girardi for the enjoyable collaboration 
on problems related to this work. 
This project received funding from the European Union's Horizon 2020 
research and innovation programme
under the Marie Sk\l{}odowska-Curie grant agreements 
No 674896 (ITN Elusives) and No 690575 (RISE InvisiblesPlus).
The work of S.T.P. was supported in part
by the INFN program on Theoretical Astroparticle Physics (TASP) 
and by the  World Premier International Research Center
Initiative (WPI Initiative, MEXT), Japan. 
A.V.T. would like to thank Kavli IPMU for the hospitality and support 
during the final stages of the work on this project.

\bibliography{ViabilityA4S4A5}

\end{document}